\documentclass{wdg_radon_series_surveys}

\usepackage{srcltx}
\title{
Optimal consumption and investment
with bounded  downside risk measures for logarithmic utility functions
}
\headlinetitle{
Optimal consumption and investment
}

\lastnameone{Kl\"uppelberg}
\firstnameone{Claudia}
\addressone{Center for Mathematical Sciences,
Technische Universit\"at M\"unchen, D-85747 Garching}
\countryone{Germany}
\emailone{cklu@ma.tum.de}

\lastnametwo{Pergamenchtchikov}
\firstnametwo{Serguei}
\addresstwo{Laboratoire de Math\'ematiques Rapha\"el Salem,
UMR 6085 CNRS, Universit\'e de Rouen,
Avenue de l'Universit\'e, BP.12,
76801 Saint Etienne du Rouvray}
\countrytwo{France}
\emailtwo{Serge.Pergamenchtchikov@univ-rouen.fr}
\thankstwo{This work was supported by the European Science Foundation
through the AMaMeF programme.}

\newcommand{\ben}{\begin{enumerate}}
\newcommand{\een}{\end{enumerate}}

\newcommand{\beq}{\begin{eqnarray}}
\newcommand{\eeq}{\end{eqnarray}\noindent}

\newcommand{\bde}{\begin{definition}}
\newcommand{\ede}{\end{definition}}

\newcommand{\btab}{\begin{tab}}
\newcommand{\etab}{\end{tab}}

\newcommand{\beao}{\begin{eqnarray*}}
\newcommand{\eeao}{\end{eqnarray*}\noindent}

\newcommand{\beam}{\begin{eqnarray}}
\newcommand{\eeam}{\end{eqnarray}\noindent}

\newcommand{\barr}{\begin{array}}
\newcommand{\earr}{\end{array}}

\newcommand{\bdis}{\begin{displaymath}}
\newcommand{\edis}{\end{displaymath}\noindent}

\def\bbr{{\mathbb R}}

\newcommand{\al}{{\alpha}}
\newcommand{\la}{{\lambda}}

\newcommand{\argmax}{{\rm argmax}}

\newcommand{\ov}{\overline}
\newcommand{\wh}{\widehat}
\newcommand{\wt}{\widetilde}

\newcommand\cE{{\cal E}}

\newcommand\cF{{\cal F}}
\newcommand\cL{{\cal L}}

\newcommand\cU{{\cal U}}
\newcommand\cV{{\cal V}}

\def\text#1{\hbox{#1}}
\def\halmos{\mbox{\hfill $\Box$}}

\def\E{{\bf E}}
\def\P{{\bf P}}

\def\W{{\bf W}}
\def\C{{\bf C}}

\def\Chi{{\bf 1}}
\def\d{\mbox{d}}
\def\build #1_#2{\mathrel{\mathop{\kern 0pt #1}\limits_{#2}}} 
\newcommand{\zs}[1]{{\mathchoice{#1}{#1}{\lower.25ex\hbox{$\scriptstyle#1$}}
{\lower0.25ex\hbox{$\scriptscriptstyle#1$}}}}

\numberwithin{equation}{chapter}

\firstpage{1}                                       %

\abstract{We investigate optimal consumption problems for a Black-Scholes market
under uniform restrictions on Value-at-Risk and Expected Shortfall for
logarithmic utility functions. We find the solutions in terms of a dynamic strategy in explicit form, which can be compared and interpreted.
This paper continues our previous work, where we solved similar problems for power utility functions.}



\begin{document}

\chapter{Introduction}\label{sec:1}

One of the principal questions in mathematical finance is the optimal investment/con\-sump\-tion problem for continuous time market models.
By applying results from sto\-chastic control theory, explicit solutions have been obtained for some special cases
(see e.g. Karatzas and Shreve \cite{KaSh1}, Korn \cite{Ko} and references therein).

With the rapid development of the derivatives markets, together with
margin tradings on certain financial products,
the exposure to losses of investments into risky assets can be considerable.
Without a careful analysis of the potential danger, the investment can cause catastrophic consequences such as, for example, the recent crisis in the ``Soci\'et\'e G\'en\'erale''.

To avoid such situations the Basel Committee on Banking Supervision in 1995
suggested some measures for the assessment of market risks.
It is widely accepted that the {\em Value-at-Risk} (VaR) is a useful summary risk measure (see, Jorion \cite{Jo} or Dowd \cite{Do}).
We recall that the VaR is the maximum expected loss over a given horizon
period at a given confidence level.
Alternatively, the {\em Expected Shortfall} (ES) or {\em Tail Condition Expectation} (TCE) measures also the expected loss given the confidence level is violated.

In order to satisfy the Basel commitee requirements, portfolios have to control the level of VaR or (the more restrictive) ES throughout the investment horizon.
This leads to stochastic control problems under restrictions on such risk measures.

Our goal in this paper is the optimal choice of a dynamic portfolio
subject to a risk limit specified in terms of VaR or ES
uniformly over horizon time interval $[0,T]$.

In Kl\"uppelberg and Pergamenshchikov in \cite{KlPe} we considered the optimal investment/consumption problem with uniform risk limits throughout the investment horizon for power utility functions. In that paper also some interpretation of VaR and ES besides an account of the relevant literature can be found. Our results in \cite{KlPe} have interesting interpretations.
We have, for instance, shown that for power utility functions with exponents less than one, the optimal constrained strategies are riskless for sufficiently small risk bounds: they recommend consumption only.
On the contrary, for the (utility bound) of a linear utility function the optimal constrained strategies recommend to invest everything into
risky assets and consume nothing.

In this paper we investigate the optimal investment/consumption problem for logarithmitic utility functions again under constraints on uniform versions of VaR and ES over the whole investment horizon $[0,T]$.
Using optimization methods in  Hilbert functional spaces, we find all optimal solutions in explicit form.
It turns out that the  optimal constrained strategies are
the unconstrained ones multiplied by some coefficient which is less then one and depends on the specific constraints.

Consequently, we can make the main recommendation:
{\em To control the  market risk throughout the investment horizon $[0,T]$
 restrict the optimal unconstrained portfolio allocation by specific multipliers
(given in explicit form in \eqref{4.13} for the VaR constraint and
in \eqref{4.37} for the ES constraint).}

Our paper is organised as follows.
In Section~\ref{sec:2} we formulate the problem.
We define the Black-Scholes model for the price processes
and present the wealth process in terms of an SDE.
We define the cost function for the logarithmic utility function and present the admissible control processes.
We also present the unconstrained consumption and investment problem of utility maximization for logarithmic utility.
In Sections~\ref{sec:4} and~\ref{subsec:4.3}  we consider the constrained problems.
Section~\ref{sec:4} is devoted to a risk bound in terms of Value-at-Risk, whereas Section~\ref{sec:5} discusses the consequences of a risk bound in terms of Expected Shortfall.
Auxiliary results and proofs are postponed to Section~\ref{sec:6}.
We start there with material needed for the proofs of both regimes, the Value-at-Risk and the ES risk bounds.
In Section~\ref{sec:5} all proofs of Section~\ref{sec:4} can be found, and in Section~\ref{sec:5} all proofs of Section~\ref{sec:5}.
Some technical lemmas postponed to the Appendix, again divided in two parts for the Value-at-Risk regime and the ES regime.

\chapter{Formulating the problem}\label{sec:2}

\section{The model and first results}\label{subsec:2.1}

We work in the same framework of self-financing portfolios as in
Kl\"uppelberg and Pergamenshchikov in \cite{KlPe}, where the financial market
is of Black-Scholes type consisting of one
 {\em riskless bond} and several   {\em risky stocks} on the interval $[0,T]$.
 Their respective
prices $S_\zs{0}=(S_\zs{0}(t))_{0\le t\le T}$ and
$S_\zs{i}=(S_\zs{i}(t))_{0\le t\le T}$ for $i=1,\ldots,d$
evolve according to the equations:
\begin{equation}\label{2.1}
\left\{\begin{array}{ll}
\d S_\zs{0}(t)\,=\,r_\zs{t}\,S_\zs{0}(t)\,\d t\,, & S_\zs{0}(0)\,=\,1\,,\\[5mm]
\d S_\zs{i}(t)\,=\,S_\zs{i}(t)\,\mu_\zs{i}(t)\,\d t\,+\,S_\zs{i}(t)\,
\sum^d_\zs{j=1}\,\sigma_\zs{ij}(t)\,\d W_\zs{j}(t)\,, &
S_\zs{i}(0)>0\,.
\end{array}\right.
\end{equation}
Here $W_\zs{t}=(W_\zs{1}(t),\ldots,W_\zs{d}(t))'$
is a standard $d$-dimensional Wiener process in $\bbr^d$;
$r_\zs{t}\in\bbr$ is the {\em riskless interest rate};
$\mu_\zs{t}=(\mu_\zs{1}(t),\ldots,\mu_\zs{d}(t))'$ is the vector of
{\em stock-appreciation rates} and
$\sigma_\zs{t}=(\sigma_\zs{ij}(t))_{1\le i,j\le d}$ is
the matrix of {\em stock-volatilities}. We assume that the coefficients
$(r_\zs{t})_\zs{0\le t\le T}$,
$(\mu_\zs{t})_\zs{0\le t\le T}$ and $(\sigma_\zs{t})_\zs{0\le t\le T}$
are deterministic cadlag functions.
We also assume that the
matrix $\sigma_\zs{t}$ is non degenerated for all $0\le t\le T$.

We denote  by $\cF_\zs{t}=\sigma\{W_\zs{s}\,,s\le t \}$, $t\ge 0$, the filtration generated by the Brownian motion (augmented by the null sets).
Furthermore, $|\cdot|$ denotes the Euclidean norm for vectors and the
corresponding matrix norm for matrices and  prime denotes the transposed.
For $(y_\zs{t})_\zs{0\le t\le T}$ square integrable over the fixed interval $[0,T]$ we define
$\|y\|_T=(\int_0^T |y_\zs{t}|^2\,\d t)^{1/2}$.

The portfolio process$\left(\pi_\zs{t}=(\pi_\zs{1}(t),\ldots, \pi_\zs{d}(t))'\right)_\zs{0\le t\le T}$ represents the
fractions of the wealth process invested into the stocks.
The consumption rate is denoted by $(v_\zs{t})_\zs{0\le t\le T}$.

Then (see~\cite{KlPe} for details) the wealth process $(X_\zs{t})_{0\le t\le T}$ is the solution to the SDE
\begin{equation}\label{2.4}
\d X_\zs{t}\,=\,X_\zs{t}\,(r_\zs{t}\,+\,y'_\zs{t}\,\theta_\zs{t}-v_\zs{t})
\,\d t\,+\,X_\zs{t}\,y'_\zs{t}\,\d W_\zs{t}\,,\quad
 X_\zs{0}\,=\,x>0\,,
\end{equation}
where
$$
\theta_\zs{t}=\sigma^{-1}_\zs{t}(\mu_\zs{t}-r_\zs{t}\,\Chi)
\quad\mbox{with}\quad \Chi=(1,\ldots,1)'\in\bbr^d\,,
$$
and we assume that
$$
\int^T_\zs{0}|\theta_\zs{t}|^2\,\d t\,
\,<\,\infty
\,.
$$
The control variables are $y_\zs{t}=\sigma'_\zs{t}\pi_\zs{t}\in\bbr^d$ and $v_\zs{t}\ge0$.
More precisely, we define the $(\cF_\zs{t})_\zs{0\le t\le T}$-progressively measurable control process as $\nu=(y_\zs{t},v_\zs{t})_{t\ge 0}$, which satisfies
\begin{equation}\label{2.7}
\int^T_0\,|\,y_\zs{t}\,|^2\,\d t\,<\,\infty
\quad\mbox{and}\quad \int^T_0\,v_\zs{t}\,\d t\,<\,\infty
\quad\mbox{a.s..}
\end{equation}
In this paper we consider logarithmitic utility functions.
Consequently, we assume throughout that
\begin{equation}\label{2.7-1}
\int_0^T (\ln v_\zs{t})_\zs{-} \d t<\infty
\quad\mbox{a.s.,}
\end{equation}
where $(a)_\zs{-}=-\min(a,0)$.

To emphasize that the wealth process \eqref{2.4} corresponds to
some control process $\nu$ we write $X^{\nu}$.
Now we describe the set of control processes.
\bde\label{De.2.1}
A stochastic control process
$\nu=(\nu_\zs{t})_{0\le t\le T}=((y_\zs{t},v_\zs{t}))_{0\le t\le T}$ is called
{\em admissible}, if it is
$(\cF_\zs{t})_{0\le t\le T}$-progressively measurable
with values in $\bbr^d\times \bbr_\zs{+}$,
satisfying  integrability conditions \eqref{2.7}--\eqref{2.7-1}
such that the SDE \eqref{2.4} has a unique strong a.s.
positive continuous solution
$(X^{\nu}_\zs{t})_{0\le t\le T}$ for which
$$
\E\left(\int_0^T\left(\ln (v_t X_t^\nu) \right)_\zs{-}\d t + \left(\ln X_T^\nu\right)_\zs{-}\right)<\infty\,.
$$
We denote by $\cV$ the class of all {\em admissible control processes}.
\ede

For $\nu\in \cV$ we define the cost function
\begin{equation}\label{2.10}
J(x,\nu):=\E_x\,\left(\int^T_0\,
\ln \left(v_\zs{t}\, X^{\nu}_\zs{t}\right)\,\d t\,+\,
\ln X^{\nu}_\zs{T}\right)\,.
\end{equation}
Here $\E_\zs{x}$ is the expectation operator conditional on  $X^{\nu}_\zs{0}=x$.

We recall a well-known result, henceforth called the {\em unconstrained problem}:
\begin{equation}\label{3.1}
\max_\zs{\nu\in\cV}\,J(x,\nu)\,.
\end{equation}
To formulate the solution we set
\beao
\omega(t)={T-t+1}
\quad\mbox{and}\quad
\wh r_\zs{t}=r_\zs{t}\,+\,\frac{|\theta_\zs{t}|^2}{2}\,,\quad 0\le t\le T\,.
\eeao

\begin{theorem}[Karatzas and Shreve~\cite{KaSh1}, Example~6.6, p.~104]\label{Th.3.1}\hspace{0cm}\\
The optimal value of $J(x,\nu)$ is given by
\beao
\max_\zs{\nu\in\cV}\,J(x,\nu)
\,=\, J(x,\nu^*)\,=\,(T+1)\,\ln\,\frac{x}{T+1}\,+\,
\int^{T}_\zs{0}\,\omega(t)\, \wh r_\zs{t}\,\d t\,.
\eeao
The optimal control process
$\nu^*=(y^*_\zs{t},v^*_\zs{t})_\zs{0\le t\le T}\in\cV$ is of the form
\begin{equation}\label{3.4}
y^*_\zs{t}\,=\,\theta_\zs{t}\quad\mbox{and}\quad
v^*_\zs{t}\,=\,\frac{1}{\omega(t)}\,,
\end{equation}
where the optimal wealth process $(X^*_\zs{t})_{0\le t\le T}$
 is given as the solution to
\begin{equation}\label{3.5}
\d X^*_\zs{t}\,=\,X^*_\zs{t}\,\left(r_\zs{t}\,+
\,|\theta_\zs{t}|^2\,-\,v^*_\zs{t}\right)\,\d t\,+
\,\,X^*_\zs{t}\,\theta'_\zs{t}\,\d W_\zs{t}\,,\quad X^*_\zs{0}\,=\,x\,,
\end{equation}
which is
\beao
X^{*}_\zs{t} & = & x\,\frac{T+1-t}{T+1}\,\exp\Big(\int^t_0\,
\wh r_\zs{u}\,\d u\,+\,
\int^t_0\,\theta'_\zs{u}\,\d W_\zs{u}\Big)\,.
\eeao
\end{theorem}

Note that the optimal solution \eqref{3.4} of problem \eqref{3.1}
is deterministic, and we denote in the following by $\cU$ the set of deterministic functions
$\nu=(y_\zs{t},v_\zs{t})_{0\le t\le T}$
satisfying conditions \eqref{2.7} and \eqref{2.7-1}.

For the above result we can state that
$$
\max_\zs{\nu\in\cV}\,J(x,\nu)=\max_\zs{\nu\in\cU}\,J(x,\nu)\,.
$$
Intuitively, it is clear that to construct  financial portfolios in the market model \eqref{2.1} the investor can invoke only information given by the coefficients $(r_\zs{t})_\zs{0\le t\le T}$,
$(\mu_\zs{t})_\zs{0\le t\le T}$ and $(\sigma_\zs{t})_\zs{0\le t\le T}$
which are  deterministic functions.

Then for $\nu\in\cU$, by It\^o's formula, equation \eqref{2.4} has solution
\beao\label{4.3}
 X^{\nu}_\zs{t}\,=\,x\,\cE_\zs{t}(y)\,e^{R_\zs{t}-V_\zs{t}+(y,\theta)_\zs{t}}\,,
\eeao
with $R_\zs{t}=\int^t_0 r_\zs{u}\d u$,
$V_\zs{t}=\int^t_0 v_\zs{u}\d u$,
$(y,\theta)_\zs{t}=\int^t_0 y'_\zs{u}\,\theta_\zs{u}\d u$
and the stochastic exponential
\beao\label{4.4}
\cE_\zs{t}(y)=
\exp\Big(\int^t_0 y'_\zs{u} \d W_\zs{u}-\frac{1}{2}
\int^t_0\,|y_u|^{2}\d u
\Big)\,.
\eeao
Therefore, for $\nu\in\cU$ the process $(X^{\nu}_\zs{t})_\zs{0\le t\le T}$ is positive, continuous and satisfies
\beao
\sup_\zs{0\le t\le T}\E\,|\ln X^{\nu}_\zs{t}|\,<\,\infty\,.
\eeao
This implies that $\cU\subset\cV$.
Moreover, for $\nu\in\cU$
we can calculate the cost function \eqref{2.10} explicitly as
\begin{align}\nonumber
J(x,\nu)=(T+1)\ln x\,+&\,\int^T_\zs{0}\omega(t) \left(r_\zs{t}+y'_\zs{t}\theta_\zs{t}-\frac{1}{2} |y_\zs{t}|^ 2\right)\d t
\\ \label{4.5-1}
&+\int^T_\zs{0}(\ln v_\zs{t}-V_\zs{t})\d t\,-V_\zs{T}\,.
\end{align}

\chapter{Optimization with constraints: main results}\label{sec:4}

\section{Value-at-Risk constraints }\label{subsec:4.2}

As in Kl\"uppelberg and Pergamenchtchikov~\cite{KlPe} we use as risk measures the modifications of  Value-at-Risk and  Expected Shortfall introduced in Emmer, Kl\"uppelberg and Korn~\cite{EmKlKo}, which reflect the capital reserve.
For simplicity, in order to avoid non-relevant cases, we consider only $0<\alpha<1/2$.

\bde\label{De.4.1} {\em [Value-at-Risk (VaR)]}\\
For a control process $\nu$ and $0<\alpha\le 1/2$ define the {\em Value-at-Risk (VaR)} by
\beao
{\rm VaR}_\zs{t}(\nu,\alpha):=x\,e^{R_\zs{t}}-Q_\zs{t}\,,
\quad t\ge 0\,,
\eeao
where for $t\ge 0$ the quantity
$Q_\zs{t}=\inf\{z\ge 0\,:\,
\P\left(X^{\nu}_\zs{t}\,\le\,z\right)\,\ge \,\alpha\}$
is the $\alpha$-quantile  of $X^{\nu}_\zs{t}$.
\ede

Note that for every $\nu\in\cU$ we find
\begin{equation}\label{4.7}
Q_\zs{t}\,=\,x\,
\exp\left(R_\zs{t}-V_\zs{t}+(y,\theta)_\zs{t}-\frac{1}{2}\|y\|^2_\zs{t}
-|q_\zs{\alpha}|\|y\|_\zs{t}\right)\,,
\end{equation}
where $q_\zs{\alpha}$ is the $\alpha$-quantile of the standard normal distribution.

We define the {\em level risk function} for some coefficient $0<\zeta<1$ as
\begin{equation}\label{4.8}
\zeta_\zs{t}\,=\,\zeta\,x\,e^{R_\zs{t}}\,,\quad t\in [0,T]\,.
\end{equation}

The coefficient $\zeta\in (0,1)$ introduces some risk aversion behaviour into the model.
In that sense it acts similarly as a utility function does.
However, $\zeta$ has a clear interpretation,
and every investor can choose and understand the influence of the risk bound $\zeta$ as a proportion of the riskless bond investment.

We consider the maximization problem for the cost function
\eqref{4.5-1} over strategies $\nu\in\cU$ for which the
Value-at-Risk is bounded by the level function
\eqref{4.8} over the interval $[0,T]$, i.e.
\begin{equation}\label{4.9}
\max_\zs{\nu\in\cU}\,J(x,\nu)\quad
\mbox{subject to}\quad\sup_\zs{0\le t\le T}\,
\frac{{\rm VaR}_\zs{t}(\nu,\alpha)}{\zeta_\zs{t}}\,\le\,1\,.
\end{equation}
To formulate the solution of this problem we define
\begin{equation}\label{4.10}
G(u,\lambda):=\int^{T}_\zs{0}\,
\frac{(\omega(t)+\lambda)^2}
{\left(\lambda|q_\zs{\alpha}|+u(\omega(t)+\lambda)\right)^2}
|\theta_\zs{t}|^2\,\d t\,,\quad u\ge 0\,,\lambda\ge 0\,.
\end{equation}
Moreover, for fixed $\la>0$ we denote by
\begin{equation}\label{4.11}
\rho(\lambda)=\inf\{u \ge 0\,:\,
G(u,\lambda)\le 1\}\,,
\end{equation}
if it exists, and set $\rho(\lambda)=+\infty$ otherwise.
For a proof of the following lemma see ~\ref{subsec:A.1}.

\begin{lemma}\label{Le.4.2}
Assume that $|q_\zs{\alpha}|> \|\theta\|_\zs{T}>0$ and
\beao
0\le \lambda\le
\lambda_\zs{\max}=
\frac{k_\zs{1}+\sqrt{k_\zs{2}
\left(q^{2}_\zs{\alpha}-\|\theta\|^{2}_\zs{T}\right)
+k^{2}_\zs{1}}}
{q^{2}_\zs{\alpha}-\|\theta\|^{2}_\zs{T}
}\,,
\eeao
where
$k_\zs{1}=\|\sqrt{\omega}\theta\|^2_\zs{T}$ and
$k_\zs{2}=\|\omega\theta\|^2_\zs{T}$.
Then the equation  $G(\cdot,\lambda)=1$ has the unique positive solution $\rho(\lambda)$. Moreover, $\rho(\lambda)<\infty$ for all $0\le \lambda\le \lambda_\zs{\max}$, and $\rho(\lambda_\zs{\max})=0$.
\end{lemma}

Now for $\lambda\ge 0$ fixed and $0\le t\le T$ we define the weight function
\begin{equation}\label{4.13}
\tau_\zs{\la}(t)=\frac{\rho(\lambda)(\omega(t)+\lambda)}
{\lambda|q_\zs{\alpha}|+\rho(\lambda)(\omega(t)+\lambda)}\,.
\end{equation}
Here we set $\tau_\zs{\la}(\cdot) \equiv 1$ for $\rho(\lambda)=+\infty$.
It is clear, that for every fixed $\la\ge 0$,
\begin{equation}\label{4.14}
0\le \tau_\zs{\la}(T)\le \tau_\zs{\la}(t)\le 1\,,\quad 0\le t\le T\,.
\end{equation}
To take the VaR constraint into account we define
\beao
\Phi(\lambda)=|q_\zs{\alpha}|\|\tau_\zs{\lambda}\theta\|_\zs{T}
+\frac{1}{2}\|\tau_\zs{\lambda} \theta \|^{2}_\zs{T}
-\|\sqrt{\tau_\zs{\lambda}}\theta\|^{2}_\zs{T}\,.
\eeao
Denote by $\Phi^{-1}$ the inverse of $\Phi$, provided it exists.
A proof of the following lemma is given in~\ref{subsec:A.1}.

\begin{lemma}\label{Le.4.2-1}
Assume that $\|\theta\|_\zs{T}>0$ and
\begin{equation}\label{4.16-1}
0<\zeta<1-e^{-|q_\zs{\alpha}|\|\theta\|_\zs{T}+\|\theta\|^2_\zs{T}/2}\,.
\end{equation}
Then for all $0\le a\le -\ln(1-\zeta)$
the inverse $\Phi^{-1}(a)$ exists.
Moreover,
$$
0\le \Phi^{-1}(a)< \lambda_\zs{\max}
\quad\mbox{for}\quad
0< a\le -\ln(1-\zeta)
$$
and $\Phi^{-1}(0)=\lambda_\zs{\max}$.
\end{lemma}

Now set
\begin{equation}\label{4.16-2}
\phi(\kappa):=\Phi^{-1}\left(\ln\frac{1-\kappa}{1-\zeta}\right)\,,\quad 0\le \kappa\le \zeta\,,
\end{equation}
and define the investment strategy
\begin{equation}\label{4.17}
\wt{y}^{\kappa}_\zs{t}:=\theta_\zs{t} \tau_{\phi(\kappa)}(t)\,,\quad 0\le t\le T\,.
\end{equation}
To introduce the optimal consumption rate  we define
\begin{equation}\label{4.18}
v^{\kappa}_t=\frac{\kappa}{T-t \kappa}
\end{equation}
and recall that for
\beao\label{4.19}
\kappa=\kappa_\zs{0}=\frac{T}{T+1}
\eeao
the function $v^\kappa_t$ coincides with the optimal unconstrained consumption rate $1/\omega(t)$ as defined in \eqref{3.4}.

It remains to fix the parameter $\kappa$.
To this end we introduce the cost function
\begin{equation}\label{4.20}
\Gamma(\kappa)=\ln(1-\kappa)+T\ln\kappa+ \int^{T}_\zs{0}\omega(t)\,|\theta_\zs{t}|^{2}\,
\left(
{\tau}_{\phi(\kappa)}(t)-\frac12 {\tau}_{\phi(\kappa)}^{2}(t)
\right)\,\d t\,.
\end{equation}
To choose the parameter $\kappa$ we  maximize $\Gamma$:
\begin{equation}\label{4.21}
\gamma=\gamma(\zeta)
=\mbox{\argmax}_\zs{0\le \kappa\le \zeta}
\,\Gamma(\kappa)\,.
\end{equation}
With this notation we can formulate the main result of this section.

\begin{theorem}\label{Th.4.3}
Assume that  $\|\theta\|_\zs{T}>0$.
Then for all $\zeta>0$ satisfying \eqref{4.16-1} and for all $0<\alpha<1/2$ for which
\begin{equation}\label{4.23}
|q_\zs{\alpha}|\ge 2\,(T+1)\,\|\theta\|_\zs{T}\,,
\end{equation}
the optimal value of $J(x,\nu)$ for  problem \eqref{4.9} is given by
\begin{equation}\label{4.24}
J(x,\nu^*)=\,A(x)\,+
\,\Gamma\left(\gamma(\zeta)\right)\,,
\end{equation}
where
\beam\label{A}
A(x)=(T+1)\ln x+\int^T_\zs{0}\omega(t)r_\zs{t}\, \d t-T\,\ln T
\eeam
and the optimal control
$\nu^*=(y^*_\zs{t},v^*_\zs{t})_\zs{0\le t\le T}$ is of the form
\begin{equation}\label{4.25}
y^*_\zs{t}=\wt{y}^{\gamma}_{\zs{t}}
\quad\mbox{and}\quad
v^*_\zs{t}=v^{\gamma}_{\zs{t}}\,.
\end{equation}
The optimal wealth process is the solution of the SDE
\beao
\d X^{*}_\zs{t}\,=\, X^{*}_\zs{t}\,(r_\zs{t}\,-\,v^*_\zs{t}\,+\,
(y^{*}_\zs{t})'\,\theta_\zs{t})
\,\d t\,+\,X^{*}_\zs{t}\,(y^{*}_\zs{t})'\,
\d W_\zs{t}\,,\quad X^*_\zs{0}\,=\,x\,,
\eeao
given by
$$
X^*_\zs{t}\,=\,
x\,\cE_\zs{t}(y^*)\,
\frac{T-\gamma(\zeta) t}{T}\,
e^{R_\zs{t}-V_\zs{t}+(y^*,\,\theta)_\zs{t}}\,,\quad 0\le t\le T\,.
$$
\end{theorem}

\medskip

The following corollary is a consequence of \eqref{4.5-1}.

\begin{corollary}\label{Co.4.3}
If $\|\theta\|_\zs{T}=0$, then
for all $0<\zeta<1$ and for all $0<\alpha<1/2$
$$
y^*_\zs{t}=0
\quad\mbox{and}\quad
v^*_\zs{t}=v^{\gamma}_{\zs{t}}
$$
with
$\gamma
=\mbox{\argmax}_\zs{0\le \kappa\le \zeta}
\left(\ln(1-\kappa)+T\ln\kappa\right)=\min(\kappa_\zs{0},\zeta)$.
 Moreover, the optimal wealth process  is the deterministic  function
$$
X^*_\zs{t}\,=\,x\,\dfrac{T-\min(\kappa_\zs{0},\zeta) \,t}{T}\,e^{R_\zs{t}}\,,\quad 0\le t\le T\,.
$$
\end{corollary}

In the next corollary we give some sufficient condition, for
which the investment  process equals zero (the optimal
strategy is riskless).
This is the first marginal case.

\begin{corollary}\label{Co.4.4}
Assume that $\|\theta\|_\zs{T}>0$ and that \eqref{4.16-1} and \eqref{4.23} hold.
Define
$$
|\theta|_\zs{\infty}=\sup_\zs{0\le t\le T}\,|\theta_\zs{t}|\,<\infty\,.
$$
If $0<\zeta<\kappa_\zs{0}$ and
\begin{equation}\label{4.27}
|q_\zs{\alpha}|\ge (1+T)\|\theta\|_\zs{T}
\left(
1+
\frac{\zeta(T+1)|\theta|^2_\zs{\infty}}{(1-\zeta)T-\zeta}
\right)\,,
\end{equation}
then $\gamma=\zeta$ and the optimal solution
$\nu^*=(y^*_\zs{t},v^*_\zs{t})_\zs{0\le t\le T}$
is of the form
$$
y^*_\zs{t}=0
\quad\mbox{and}\quad
v^*_\zs{t}=v^{\zeta}_\zs{t}\,.
$$
Moreover, the optimal wealth process is the deterministic function
$$
X^*_\zs{t}\,=\,x\,\dfrac{T-\zeta t}{T}\,e^{R_\zs{t}}\,,\quad 0\le t\le T\,.
$$
\end{corollary}

Below we give some sufficient conditions, for which the solution of
optimization problem \eqref{4.9} coincides with the unconstrained solution
\eqref{3.4}.
This is the second marginal case.

\begin{theorem}\label{Th.4.5}
Assume that
\begin{equation}\label{4.28}
\zeta>1-\frac{1}{T}\,e^{-|q_\zs{\alpha}|\|\theta\|_\zs{T}+\|\theta\|^2_\zs{T}/2}\,.
\end{equation}
Then for all $0<\alpha<1/2$ for which $|q_\zs{\alpha}|\ge\|\theta\|_\zs{T}$,
the solution of the optimization problem \eqref{4.9} is given by
\eqref{3.4}--\eqref{3.5}.
\end{theorem}

\section{Expected Shortfall Constraints}\label{subsec:4.3}

Our next risk measure is an analogous modification of the  {\em Expected Shortfall} (ES).

\bde {\em [Expected Shortfall (ES)]}\\
For a control process $\nu$ and $0<\alpha\le 1/2$ define
\begin{align*}
m_\zs{t}(\nu,\alpha)\,=\,
\E_\zs{x}\left(X^\nu_\zs{t}\,|\,X^\nu_\zs{t}\le\,Q_\zs{t}\right)\,,
\quad t\ge0\,,
\end{align*}
where $Q_\zs{t}$ is the $\alpha$-quantile
of $X^{\nu}_\zs{t}$ given by \eqref{4.7}.
The {\em Expected Shortfall (ES)} is then defined as
$$
ES_\zs{t}(\nu,\alpha)
\,=\,xe^{R_\zs{t}}\,-\,m_\zs{t}(\nu,\alpha)\,,\quad t\ge 0\,.
$$
\ede

Again for $\nu\in \cU$ we find
\begin{align*}
m_\zs{t}(\nu,\alpha)\,=\,x\,
F_\zs{\alpha}\,(|q_\zs{\alpha}|+\|y\|_\zs{t})\,
e^{R_\zs{t}-V_\zs{t}+(y,\theta)_\zs{t}}\,,
\end{align*}
where
\beam\label{Falpha}
F_\zs{\alpha}(z)\,=\,
\frac{1}{\int^{\infty}_\zs{|q_\zs{\alpha}|}\,e^{-t^2/2}\,\d t}\,
\int^{\infty}_\zs{z}\,e^{-t^2/2}\,\d t\,.
\eeam

We consider the maximization problem for the cost function \eqref{2.10}
over strategies $\nu\in\cU$ for which the
Expected Shortfall is bounded by the level function
\eqref{4.8} over the interval $[0,T]$, i.e.
\begin{equation}\label{4.30}
\max_\zs{\nu\in\cU}\,J(x,\nu)\quad
\mbox{subject to}\quad\sup_\zs{0\,\le\,t\,\le\,T}\,
\frac{ES_\zs{t}(\nu,\alpha)}
{\zeta_\zs{t}}\,\le\,1\,.
\end{equation}
We proceed similarly as for the VaR-coinstraint problem \eqref{4.9}.
Define
\begin{equation}\label{4.31}
G_\zs{1}(u,\lambda):=\int^{T}_\zs{0}\,
\frac{(\omega(t)+\lambda)^2}
{\left(\lambda\,\iota_\zs{\alpha}(u)+u(\omega(t)+\lambda)\right)^2}
|\theta_\zs{t}|^2\,\d t\,,\quad u\ge 0,\la\ge 0\,.
\end{equation}
where
\begin{equation}\label{4.32}
\iota_\zs{\alpha}(u)
=\frac{1}{\varpi(u+|q_\zs{\alpha}|)}-u
\quad\mbox{with}\quad
\varpi(y)=e^{\frac{y^2}{2}}\,\int^{\infty}_\zs{y}\,e^{-\frac{t^2}{2}}\,\d t
\,.
\end{equation}
It is well-known and easy to prove that
\begin{equation}\label{4.33}
\frac{1}{y}-\frac{1}{y^3}\,
\le\,
\varpi(y)\,
\le\,
\frac{1}{y}\,,\quad y>0\,.
\end{equation}
This means that $\iota_\zs{\alpha}(u)\,\ge\,|q_\zs{\alpha}|$ for all $u\ge 0$,
which implies for every fixed $\la\ge 0$ that $G_\zs{1}(u,\la)\le G(u,\la)$ for all $u\ge 0$.
Moreover, similarly to \eqref{4.11} we define
\begin{equation}\label{4.35}
\rho_\zs{1}(\lambda)=\inf\{u \ge 0\,:\, G_\zs{1}(u,\lambda)\le 1\}\,.
\end{equation}
Since $H$ has similar behaviour as $G$, the following lemma is a modification of Lemma~\ref{Le.4.2}.
Its proof is analogous to the proof of Lemma~\ref{Le.4.2}.

\begin{lemma}\label{Le.4.7}
Assume that $|q_\zs{\alpha}|> \|\theta\|_\zs{T}>0$ and
\beao
0\le \lambda\le \lambda^\prime_\zs{\max}=
\frac{k_\zs{1}+\sqrt{k_\zs{2}
\left(\psi^{2}_\zs{\alpha}(0) -\|\theta\|^{2}_\zs{T}\right)
+k^{2}_\zs{1}}}
{\psi^{2}_\zs{\alpha}(0) -\|\theta\|^{2}_\zs{T}}\,,
\eeao
where
$k_\zs{1}$ and
$k_\zs{2}$ are given in
Lemma~\ref{Le.4.2}.
Then the equation $G_\zs{1}(\cdot,\lambda)=1$ has the unique positive solution $\rho_\zs{1}(\lambda)$.
Moreover, $\rho_\zs{1}(\lambda)<\infty$ for $0\le \lambda\le \lambda^\prime_\zs{\max}$ and $\rho_\zs{1}(\lambda^\prime_\zs{\max})=0$.
\end{lemma}

Now for $\lambda\ge 0$ fixed and  $0\le t\le T$ we define the weight function
\begin{equation}\label{4.37}
\varsigma_\la(t)=\frac{\rho_\zs{1}(\lambda)\,(\omega(t)+\lambda)}
{\lambda\,\iota_\zs{\alpha}(\rho_\zs{1}(\lambda)) +\rho_\zs{1}(\lambda)\,(\omega(t)+\lambda)}\,,
\end{equation}
and we set $\varsigma_\la(\cdot)\equiv 1$ for $\rho_\zs{1}(\lambda)=+\infty$.
Note that for every fixed $\lambda\ge 0$,
\begin{equation}\label{4.38}
0\le \varsigma_\la(T)\le \varsigma_\la(t)\le 1\,,\quad 0\le t\le T\,.
\end{equation}
To take the ES constraint into account we define
\begin{equation}\label{4.39}
\Phi_\zs{1}(\lambda)\,=\,
-\|\sqrt{\varsigma_\zs{\lambda}}\theta\|^2_\zs{T}
-\ln F_\zs{\alpha}
\left(
|q_\zs{\alpha}|+ \|\varsigma_\zs{\lambda}\theta\|_\zs{T}
\right)\,.
\end{equation}
Denote by $\Phi_\zs{1}^{-1}$ the inverse of $\Phi_\zs{1}$ provided it exists.
The proof of the next lemma is given in Section~\ref{subsec:A.2}.

\begin{lemma}\label{Le.4.7-1}
Assume that $\|\theta\|_\zs{T}>0$ and
\begin{equation}\label{4.40-1}
0<\zeta<1-F_\zs{\alpha}
\left(
|q_\zs{\alpha}|+\|\theta\|_\zs{T}
\right)\,e^{\|\theta\|^2_\zs{T}}\,.
\end{equation}
Then for all
$0\le a\le -\ln(1-\zeta)$ the inverse $\Phi_\zs{1}^{-1}$ exists and
$0\le \Phi_\zs{1}^{-1}(a)< \lambda_\zs{\max}$ for $0< a\le -\ln(1-\zeta)$
and $\Phi_\zs{1}^{-1}(0)=\lambda^\prime_\zs{\max}$.
\end{lemma}

Now, similarly to \eqref{4.11} we set
\begin{equation}\label{4.41}
\phi_\zs{1}(\kappa)=\Phi_\zs{1}^{-1}\left(\ln\frac{1-\kappa}{1-\zeta}\right)\,,\quad 0\le\kappa\le\zeta\,,
\end{equation}
and define the investment strategy
\beam\label{3.30a}
\wt y_\zs{t}^{1,\kappa} = \theta_t \varsigma_\zs{\phi_\zs{1}(\kappa)}(t)\,,\quad 0\le t\le T\,.
\eeam
We introduce the cost function
\begin{equation}\label{4.42}
\Gamma_\zs{1}(\kappa)= \ln(1-\kappa)+T\ln\kappa+\int^{T}_\zs{0}\omega(t)\,|\theta_\zs{t}|^{2}\,
\left(
{\varsigma}_\zs{\phi_\zs{1}(\kappa)}(t)-\frac{1}{2} {\varsigma}^{2}_\zs{\phi_\zs{1}(\kappa)}(t)
\right)\,\d t\,.
\end{equation}
To fix the parameter $\kappa$ we maximize $\Gamma_\zs{1}$:
\begin{equation}\label{4.43}
\gamma_\zs{1}=\gamma_\zs{1}(\zeta)
=\mbox{\argmax}_\zs{0\le \kappa\le \zeta}
\,\Gamma_\zs{1}(\kappa)\,.
\end{equation}
With this notation we can formulate the main result of this section.

\begin{theorem}\label{Th.4.8}
Assume that  $\|\theta\|_\zs{T}>0$. Then
 for all $\zeta>0$ satisfying \eqref{4.40-1} and for all $0<\alpha<1/2$
satisfying
\begin{equation}\label{4.45}
|q_\zs{\alpha}|\ge \max(1, 2(T+1)\|\theta\|_T)
\end{equation}
the optimal value of $J(x,\nu)$ for the optimization problem \eqref{4.30} is given by
\beao
J(x,\nu^*)=\,A(x)\,+
\,\Gamma_\zs{1}\left(\gamma_\zs{1}(\zeta)\right)\,,
\eeao
where the function  $A$ is defined in \eqref{A}
and the optimal control
$\nu^*=(y^*_\zs{t},v^*_\zs{t})_\zs{0\le t\le T}$ is of the form
(recall the definition of $v_\zs{t}^{\kappa}$ in \eqref{4.18})
\begin{equation}\label{4.47}
y^*_\zs{t}=\wt y^{1,\gamma_\zs{1}}_\zs{t}
\quad\mbox{and}\quad
v^*_\zs{t}=v_\zs{t}^{\gamma_\zs{1}}\,.
\end{equation}
The optimal wealth process is the solution to the SDE
\beao\label{4.48}
\d X^{*}_\zs{t}\,=\, X^{*}_\zs{t}\,(r_\zs{t}\,-\,v^*_\zs{t}\,+\,
(y^{*}_\zs{t})'\,\theta_\zs{t})
\,\d t\,+\,X^{*}_\zs{t}\,(y^{*}_\zs{t})'\,
\d W_\zs{t}\,,\quad X^{*}_\zs{0}\,=\,x\,,
\eeao
given by
$$
X^*_\zs{t}\,=\,
x\,\cE_\zs{t}(y^*)\,
\frac{T-\gamma_\zs{1}(\zeta) t}{T}\,
e^{R_\zs{t}-V_\zs{t}+(y^*,\,\theta)_\zs{t}}\,,\quad 0\le t\le T\,.
$$
\end{theorem}

\begin{corollary}\label{Co.4.8}
If $\|\theta\|_\zs{T}=0$, then the optimal solution of problem
\eqref{4.30} is given in Corollary~\ref{Co.4.3}.
\end{corollary}

Similarly to the optimization problem with VaR constraint we observe two marginal cases.
Note that the following corollary is again a consequence of \eqref{4.5-1}.

\begin{corollary}\label{Co.4.9}
Assume that $\|\theta\|_T>0$ and that \eqref{4.40-1} and \eqref{4.45} hold.
Then $\gamma_\zs{1}=\zeta$ and the assertions of Corollary~\ref{Co.4.4} hold with $\kappa_1$ replaced by $\ov\kappa_1$.
\end{corollary}

\begin{theorem}\label{Th.4.10}
Assume that \marginpar{$\|\theta\|^2_{T}/2$? NO!!!}
\begin{equation}\label{4.49}
\zeta>1-\frac{1}{T+1}\,
F_\zs{\alpha}\left(
|q_\zs{\alpha}|+\|\theta\|_\zs{T}
\right)\,e^{\|\theta\|^2_\zs{T}}\,.
\end{equation}
Then
 for all $0<\alpha<1/2$ for which $|q_\zs{\alpha}|>\max(1,\|\theta\|_\zs{T})$
the solution of problem \eqref{4.30} is given by \eqref{3.4}--\eqref{3.5}.
\end{theorem}

\section{Conclusion}\label{subsec:4.4}

If we compare the optimal solutions \eqref{4.25} and \eqref{4.47} with the unconstrained optimal strategy
\eqref{3.4}, then the risk bounds forces investors to restrict their investment into the risk assets
by multiplying the unconstrained optimal strategy by the
coefficients given in \eqref{4.17} and \eqref{4.21} for VaR constraints
and \eqref{4.41} and \eqref{4.43} for ES constraints.
The impact of the risk measure constraints
enter into the portfolio process through the risk level $\zeta$
and the confidence level $\alpha$.

\chapter{Auxiliary results and proofs}\label{sec:6}

In this section we consider maximization problems
with constraints for the two terms of \eqref{4.5-1}:
\beam\label{eq4.0}
I(V):=\int_0^T (\ln v_t-V_t)\d t
\quad\mbox{and}\quad
H(y):=\int_0^T \omega(t)\left(y_t'\theta_t-\frac12 |y_t|^2\right) \d t\,.
\eeam

We start with a result concerning the optimization of $I(\cdot)$,
which will be needed to prove results from both Sections~\ref{subsec:4.2} and~\ref{subsec:4.3}.

Let $\W[0,T]$ be the set of differentiable functions $f\,:\,[0,T]\to\bbr$ having positive cadlag derivative $\dot f$ satisfying condition \eqref{2.7-1}.
For $b>0$ we define
\begin{equation}\label{A.1}
\W_\zs{0,b}[0,T]=\{f\in \W[0,T]\,:\,f(0)=0
\quad\mbox{and}\quad f(T)=b
\}\,.
\end{equation}

\begin{lemma}\label{Le.A.1}
Consider the optimization problem
\begin{equation}\label{A.2}
\max_\zs{f\in\W_\zs{0,b}[0,T]}\,I(f)\,.
\end{equation}
The optimal value of $I$ is given by
\begin{equation}\label{A.3}
I^*(b)=
\max_\zs{f\in\W_\zs{0,b}[0,T]}\,I(f)\,
=I(f^{*})=\,-T\ln T\,-\,
T\,\ln\,\frac{e^b}{e^{b}-1}\,,
\end{equation}
with optimal solution
\begin{equation}\label{A.4}
f^{*}(t)=\ln \frac{Te^{b}}{Te^{b}-t(e^{b}-1)}\,,\quad 0\le t\le T\,.
\end{equation}
\end{lemma}

\noindent{\bf Proof. \,}
Firstly, we consider the optimization problem \eqref{A.2} in the space $\C^2[0,T]$ of two times continuously differentiable functions on $[0,T]$:
$$
\max_\zs{f\in\W_\zs{0,b}[0,T]\cap \C^2[0,T]}\,I(f)\,,
$$
By variational calculus
methods we find that it has solution \eqref{A.3}; i.e.
$$
\max_\zs{f\in\W_\zs{0,b}[0,T]\cap \C^2[0,T]}\,I(f)\,
=I(f^{*})\,,
$$
where the optimal solution $f^{*}$ is given in \eqref{A.4}.\\
Take now $f\in\W_\zs{0,b}[0,T]$ and suppose first that its derivative
$$
\dot{f}_\zs{\min}=\inf_\zs{0\le t\le T}\,\dot{f}(t)\,>0\,.
$$
Let  $\Upsilon$ be a positive two times differentiable function on $[-1,1]$
such that $\int^{1}_{-1}\,\Upsilon(z)\,\d z=1$,
and set $\Upsilon(z):=0$ for $|z|\ge 1$.
We can take, for example,
$$
\Upsilon(z)=
\left\{\begin{array}{ll}
\frac1{\int^{1}_\zs{-1}\exp\left(-\frac{1}{1-\upsilon^{2}}\right)\d \upsilon}\,\exp\left(-\frac{1}{1-z^{2}}\right)
& \quad\mbox{if}\quad |z|\le 1\,,\\[5mm]
0 & \quad\mbox{if}\quad |z|> 1\,.
\end{array}\right.
$$
By setting $\dot{f}(t)=\dot{f}(0)$ for all $t\le 0$
and $\dot{f}(t)=\dot{f}(T)$ for all $t\ge T$,
we define an approximating sequence of functions by
$$
\upsilon_\zs{n}(t)=n\int_\zs{\bbr}\,\Upsilon(n(u-t))\,\dot{f}(u)\,\d u\,.
$$
It is clear that  $(\upsilon_\zs{n})_\zs{n\ge 1}\in \C^{2}[0,T]$. Moreover,
we recall that $\dot{f}$ is cadlag, which implies that
it is bounded on  $[0,T]$; i.e.
$$
\sup_\zs{0\le t\le T}\,\dot{f}(t):=\dot{f}_\zs{\max}\,<\,\infty\,,
$$
and its discontinuity set has Lebesgue measure zero.
Therefore, the sequence $(\upsilon_\zs{n})_\zs{n\ge 1}$ is bounded; more preceisly,
\begin{equation}\label{A.6}
0<\dot{f}_\zs{\min}
\le\,
\upsilon_\zs{n}(t)\,
\le \dot{f}_\zs{\max}\,<\,\infty\,,\quad 0\le t\le T\,,
\end{equation}
and $\upsilon_\zs{n}\to \dot{f}$  as $n\to\infty$ for Lebesgue almost all  $t\in[0,T]$.
Therefore, by the Lebesgue convergence theorem we obtain
$$
\lim_\zs{n\to\infty}\,
\int^{T}_\zs{0}\,|\upsilon_\zs{n}(t)-\dot{f}(t)|\,\d t\,=\,0\,.
$$
Moreover, inequalities \eqref{A.6} imply
$$
|\ln \upsilon_\zs{n}|\,\le \ln \left(\max(\dot{f}_\zs{\max}\,,\,1)\right)+
|\ln \left(\min(\dot{f}_\zs{\min}\,,\,1)\right) | \,.
$$
Therefore,  $f_\zs{n}(t)=\int^t_\zs{0}\,\upsilon_\zs{n}(u)\,\d u$
belongs to $\Gamma_{\zs{b_\zs{n}}}\cap \C^{2}[0,T]$ for $b_n:=\int_0^T \upsilon_\zs{n}(u)\,\d u$.
It is clear that
$$
\lim_\zs{n\to\infty}\,I(f_\zs{n})\,=\,I(f)
\quad\mbox{and}\quad
\lim_\zs{n\to\infty} b_n=b\,.
$$
This implies that
$$
I(f)\le I^{*}(b)\,,
$$
where $I^{*}(b)$ is defined in \eqref{A.3}.\\
Consider now the case, where $\inf_\zs{0\le t\le T}\,\dot{f}(t)=0$.
For $0<\delta<1$ we consider the approximation sequence of functions
$$
\wt{f}_\zs{\delta}(t)=\max(\delta\,,\,\dot{f}(t))
\quad\mbox{and}\quad
f_\zs{\delta}(t)=\int^{t}_\zs{0}\,\wt{f}_\zs{\delta}(u)\,\d u\,,\quad 0\le t\le T\,.
$$
It is clear that
$f_\zs{\delta}\in\Gamma_\zs{b_\zs{\delta}}$ for
$b_\zs{\delta}=\int^T_\zs{0}\,\wt{f}_\zs{\delta}(t)\,\d t$.
Therefore, $I(f_\zs{\delta})\le I^*(b_\zs{\delta})$.
Moreover, in view of the convergence
$$
\lim_\zs{\delta\to 0}
\int^T_\zs{0}\left(\wt{f}_\zs{\delta}(t)-\dot{f}(t)\right)\d t=0
$$
we get $\limsup_\zs{\delta\to 0} I(f_\zs{\delta})\le I^*(b)$.
Moreover, note that
\begin{align*}
|I(f_\zs{\delta})-I(f)|&\le
\int_\zs{A_\zs{\delta}}\,
(\ln \delta-\ln \dot{f}(t))\,\d t\,+\,
T\,\int_\zs{A_\zs{\delta}}\,\left(
\delta-\dot{f}(t)
\right)\,\d t\\
&\le
\,\int_\zs{A_\zs{\delta}}\,(\ln \dot{f}(t))_\zs{-}\,\d t\,+
\,\delta\,
T\,
\Lambda(A_\zs{\delta})\,,
\end{align*}
where
$A_\zs{\delta}=\{t\in [0,T]\,:\,0\le\dot{f}(t)\le \delta\}$
and
$\Lambda(A_\zs{\delta})$ is the Lebesgue measuere of $A_\zs{\delta}$.
Moreover, by the definition of the $\W[0,T]$ in \eqref{A.1} the Lebesgue measure of the set
$\{t\in [0,T]\,:\,\dot{f}(t)=0\}$ equals to zero
and
$\int_0^T(\ln \dot{f}_\zs{t})_\zs{-}\,\d t<\infty$.
This implies that
$\lim_{\delta\to 0} \Lambda(A_\zs{\delta})=0$ and hence
$$
\lim_\zs{\delta\to 0}\,I(f_\zs{\delta})=I(f)\,,
$$
i.e. $I(f)\le I^*(b)$.
\halmos

\medskip

In order to deal with $H$ as defined in \eqref{eq4.0} we need some preliminary result.
As usual, we denote by $\cL_\zs{2}[0,T]$ the Hilbert space of functions $y$ satisfying the square integrability condition in \eqref{2.7}.

Define for $y\in\cL_\zs{2}[0,T]$ with $\|y\|_\zs{T}>0$
\begin{equation}\label{5.7}
\ov{y}_\zs{t}=y_\zs{t}/\|y\|_\zs{T}
\quad\mbox{and}\quad
l_\zs{y}(h)=\|y+h\|_\zs{T}-
\|y\|_\zs{T}-(\ov{y},h)_\zs{T}\,.
\end{equation}

We shall need the following lemma.

\begin{lemma}\label{Le.A.2}
Assume that $y\in \cL_\zs{2}[0,T]$ and $\|y\|_\zs{T}>0$.
Then for every $h\in\cL_\zs{2}[0,T]$ the function $l_\zs{y}(h)\ge0$.
\end{lemma}

\noindent{\bf Proof. \,}
Obviously, if $h\equiv ay$ for some $a\in\bbr$, then $l_\zs{y}(h)=(|1+a|-1-a)\|y\|_\zs{T}\ge 0$.
Let now $h \not\equiv ay$ for all $a\in\bbr$. Then
$$
l_\zs{y}(h)=\frac{2(y\,,\,h)_\zs{T}+\|h\|^2_\zs{T}}{\|y+h\|_\zs{T}+\|y\|_\zs{T}}
-(\ov{y},h)_\zs{T}
=\frac{\|h\|^2_\zs{T}-(\ov{y},h)_\zs{T}((\ov{y},h)_\zs{T}+l_\zs{y}(h) )}
{\|y+h\|_\zs{T}+\|y\|_\zs{T}}\,.
$$
It is easy to show directly that for all $h$
$$
\|y+h\|_\zs{T}+\|y\|_\zs{T}+(\ov{y},h)_\zs{T}\ge 0
$$
with equality if and only if $h\equiv ay$ for some $a\le -1$.\\
Therefore, if $h\not\equiv ay$,  we obtain
$$
l_\zs{y}(h)=
\frac{\|h\|^2_\zs{T}-(\ov{y}\,,\,h)^2_\zs{T}}{\|y+h\|_\zs{T}+\|y\|_\zs{T}+(\ov{y},h)_\zs{T}}\,\ge\,0\,.
\hfill\eqno{\Box}
$$

\section{Results and proofs of Section~\ref{subsec:4.2}}\label{sec:5}

We introduce the constraint $K:\,\cL_\zs{2}[0,T]\to \bbr$ as
\begin{equation}\label{5.1}
K(y):=\frac{1}{2}\|y\|_\zs{T}^2+
|q_\zs{\alpha}|\,\|y\|_\zs{T} -(y,\theta)_\zs{T}
\end{equation}
For  $0<a\le -\ln (1-\zeta)$ we consider the following optimization problems
\begin{equation}\label{5.3}
\max_\zs{y\in \cL_\zs{2}[0,T]}\,H(y)
\quad\mbox{subject to}\quad
K(y)=a
\end{equation}

\begin{proposition}\label{prop1}
Assume that the conditions of Lemma~\ref{Le.4.2-1} hold.
Then the optimization problem \eqref{5.3} has the unique solution
$y^*=\wt y^a=\theta_t\tau_{\lambda_\zs{a}}(t)$ with $\lambda_\zs{a}=\Phi^{-1}(a)$.
\end{proposition}

\noindent{\bf Proof. \,}
According to Lagrange's method we consider the following unconstrained
problem
\begin{equation}\label{5.5}
\max_\zs{y\in \cL_\zs{2}[0,T]}\,\Psi(y,\lambda)\,,
\end{equation}
where $\Psi(y,\lambda)=H(y)-\lambda K(y)$
and $\lambda\in\bbr$ is the Lagrange multiplier.
Now it suffices
to find some $\lambda\in\bbr$ for which the problem \eqref{5.5}
has a solution, which satisfies the constraint in \eqref{5.3}.
To this end we  represent $\Psi$  as
$$
\Psi(y,\lambda)=\int^{T}_\zs{0}
\left(
\omega(t)+\lambda
\right)\,
\left(
y'_\zs{t}\theta_\zs{t}-\frac12 |y_\zs{t}|^2
\right)
\,\d t\,-\,\lambda\,|q_\zs{\alpha}|\,
\|y\|_\zs{T}\,.
$$
It is easy to see that for $\lambda<0$ the maximum in
\eqref{5.5} equals $+\infty$; i.e. the problem \eqref{5.3}
has no solution.
Therefore, we assume that $\lambda\ge 0$.
First we  calculate the Fr\'echet derivative; i.e.
the linear operator $D_\zs{y}(\cdot,\lambda):\cL_\zs{2}[0,T]\to\bbr$
defined for $h\in\cL_\zs{2}[0,T]$ as
$$
D_\zs{y}(h,\lambda)=\lim_\zs{\delta\to 0}\,
\frac{\Psi(y+\delta h,\lambda)-\Psi(y,\lambda)}{\delta}\,.
$$
For $\|y\|_\zs{T}>0$ we obtain
$$
D_\zs{y}(h,\lambda)=\int^{T}_\zs{0}\,
(d_\zs{y}(t,\lambda))'h_\zs{t}\,\d t
$$
with
$$
d_\zs{y}(t,\lambda)=(\omega(t)+\lambda)(\theta_\zs{t}-y_\zs{t})
-\lambda |q_\zs{\alpha}|\,\ov{y}_\zs{t}\,.
$$
If $\|y\|_\zs{T}=0$, then
$$
D_\zs{y}(h,\lambda)=\int^{T}_\zs{0}\,
(\omega(t)+\lambda)\,\theta'_\zs{t}\,h_\zs{t}\,\d t
-\lambda |q_\zs{\alpha}|\,\|h\|_\zs{T}\,.
$$
Define now
\begin{equation}\label{5.6}
\Delta_\zs{y}(h,\lambda)=\Psi(y+h,\lambda)-\Psi(y,\lambda)-D_\zs{y}(h,\lambda)\,.
\end{equation}
We have to show that $\Delta_\zs{y}(h,\lambda\le 0$  for all $y,h\in\cL_\zs{2}[0,T]$.
Indeed, if $\|y\|_\zs{T}=0$ then
$$
\Delta_\zs{y}(h,\lambda)=-\frac{1}{2}\,\int^{T}_\zs{0}\,(\omega(t)+\lambda)\,
|h_\zs{t}|^2\,\d t\,\le\,0\,.
$$
If $\|y\|_\zs{T}>0$, then
$$
\Delta_\zs{y}(h,\lambda)=-\frac{1}{2}\,\int^{T}_\zs{0}\,(\omega(t)+\lambda)\,
|h_\zs{t}|^2\,\d t\,-\lambda\,|q_\zs{\alpha}|\,l_\zs{y}(h)\le 0\,,
$$
by Lemma~\ref{Le.A.2} for all $\lambda\ge 0$ and for all $y,h\in\cL_\zs{2}[0,T]$. \\
To find the solution of the optimization problem \eqref{5.5}  we have to find
$y\in\cL_\zs{2}[0,T]$ such that
\begin{equation}\label{5.8}
D_\zs{y}(h,\lambda)=0
\quad\mbox{ for all }\quad
h\in\cL_\zs{2}[0,T]\,.
\end{equation}
First notice that for $\|\theta\|_\zs{T}>0$, the solution of \eqref{5.8} can not be zero, since
for $y=0$ we obtain $D_\zs{y}(h,\lambda)<0$ for $h=-\theta$.
Consequently, we have to find an optimal solution to \eqref{5.8} for $y$ satisfying $\|y\|_\zs{T}>0$.
This means we have to find a non-zero  $y\in\cL_\zs{2}[0,T]$ such that
$$
d_\zs{y}(t,\lambda)=0\,.
$$
One can show directly that for
$0\le \lambda\le \lambda_\zs{\max}$
the unique solution of this equation is given by
\begin{equation}\label{5.8-1}
y^{\lambda}_\zs{t}:=\theta_\zs{t}\tau_\zs{\la}(t)\,,
\end{equation}
where $\tau_\la(t)$ is defined in  \eqref{4.13}.
It remains to  choose the Lagrage multiplier $\lambda$ so that it satisfies the constraint in \eqref{5.3}.
To this end note that
$$
K(y^\lambda)=\Phi(\lambda)\,.
$$
Under the conditions of Lemma~\ref{Le.4.2-1} the inverse of $\Phi$ exists.
Thus the function $y^{\lambda_\zs{a}}\not\equiv 0$
with $\lambda_\zs{a}=\Phi^{-1}(a)$
 is the solution of the problem \eqref{5.3}.
\halmos

\medskip

We are now ready to proof the main results in Section~\ref{subsec:4.2}.
The auxiliary lemmas are proved in \ref{subsec:A.1}.

\medskip

\noindent{\bf Proof of Theorem~\ref{Th.4.3}. \,}
In view of the representation of the cost function $J(x,\nu)$ in the form \eqref{4.5-1}, \marginpar{???}
we start to maximize $J(x,\nu)$ by maximizing $I$ over all functions $V$.
To this end we fix the last value of the consumption process, by setting
$\kappa=1-e^{-V_\zs{T}}$. By Lemma~\ref{Le.A.1} we find that
$$
I(V)\le I(V^{\kappa})
= -T\ln T\,+\,T\ln \kappa\,,
$$
where
\begin{equation}\label{6.1}
V^{\kappa}_t=\int^t_\zs{0}v^{\kappa}(t)\d t=
\ln \frac{T}{T-\kappa t}\,,\quad 0\le t\le T\,.
\end{equation}
Define now
$$
L_\zs{t}(\nu)\,=\,(y,\theta)_\zs{t}\,-\,\frac{1}{2}\|y\|^2_\zs{t}-V_\zs{t}-
|q_\zs{\alpha}|\,\|y\|_\zs{t}\,,\quad 0\le t\le T\,,
$$
and note that condition \eqref{4.9} is equivalent to
\begin{equation}\label{6.2}
\inf_\zs{0\le t\le T}\,L_\zs{t}(\nu)\,\ge\,
\ln\,\left(1-\zeta\right)\,.
\end{equation}
Firstly, we consider the bound in \eqref{6.2} only at time $t=T$:
$$
L_\zs{T}(\nu)\,\ge\,
\ln\,\left(1-\zeta\right)\,.
$$
Recall definition \eqref{5.1} of $K$ and choose the function $V$
as $V^\kappa$ as in \eqref{6.1}.
Then we can rewrite the bound for $L_T(\nu)$ as a bound for $K$ and obtain
$$
K(y)\,\le\,\ln\frac{1-\kappa}{1-\zeta}\,,\quad 0\le \kappa\le \zeta\,.
$$
To find the optimal investment strategy we need to solve the optimization problem \eqref{5.3} for $0\le a\le \ln(({1-\kappa})/({1-\zeta}))$.
By Proposition~\ref{prop1}  for $0<a\le -\ln(1-\zeta)$
\begin{equation}\label{6.3}
\max_\zs{y\in \cL_\zs{2}[0,T]\,,\,K(y)=a}\,H(y)=H(\wt{y}^a):=C(a)\,,
\end{equation}
where the solution $\wt{y}^a$ is defined in Proposition~\ref{prop1}.
Note that the definitions of the functions $H$ and $\wt{y}^a$
imply
$$
C(a)=\int^{T}_\zs{0}\,\omega(t)\left(
\tau_{\la_\zs{a}}(t)-\frac{1}{2}\tau_{\la_\zs{a}}^2(t)
\right)\,
|\theta_\zs{t}|^2
\d t
\quad\mbox{with}\quad
\la_\zs{a}=\Phi^{-1}(a)
\,.
$$

To consider the optimization problem \eqref{5.3} for $a=0$ we observe that
$$
K(y)
\ge
\|y\|_\zs{T}\,\left(
|q_\zs{\alpha}| -\|\theta\|_\zs{T}
\right)+
\frac{1}{2}\|y\|^2_\zs{T}\ge0\,,
$$
provided that $|q_\zs{\alpha}|>\|\theta\|_\zs{T}$
(which follows from  \eqref{4.23}).
Thus, there exists only one function for which
$K(y)=0$, namely $y\equiv 0$.
Furthermore, by Lemma~\ref{Le.4.2} $\rho(\lambda_\zs{\max})=0$ and,
therefore, definition \eqref{4.13} implies
\begin{equation}\label{6.4}
\tau_{\lambda_\zs{\max}}(\cdot)\equiv 0\,,
\quad
y^{\lambda_\zs{\max}}\equiv 0
\quad\mbox{and}\quad
 \Phi(\lambda_\zs{\max})= 0\,.
\end{equation}
This means that $\lambda_\zs{max}=\Phi^{-1}(0)$ and
$y^{\Phi^{-1}(0)}=0$;
i.e. $y^{\la_\zs{a}}$ with $\la_\zs{a}=\Phi^{-1}(a)$
is the solution of
the optimization problem  \eqref{5.3} for all $0\le a\le -\ln(1-\zeta)$.
Now we calculate the derivative of $C(a)$:
$$
\frac{\d }{\d a} C(a)=
\dot{\la}_\zs{a}\,
\int^{T}_\zs{0}\,\omega(t)\left(
1-\tau_{\la_\zs{a}}(t)
\right)\,
|\theta_\zs{t}|^2\,
\left(
\frac{\partial{\tau_{\la}(t)}}{\partial{\lambda}}|_{\la=\la_\zs{a}}\,
\right)
\d t\,,
$$
Since
$\dot{\la}_\zs{a} =1/\dot{\Phi}(\la_\zs{a})$,
by Lemma~\ref{Le.A.3}, the derivative of $C(a)$ is positive.
Therefore,
$$
\max_\zs{0\le a\le \ln((1-\kappa)/(1-\zeta))}\,C(a)\,
=\,C\left(\ln\frac{1-\kappa}{1-\zeta}\right),
$$
and we choose $a=\ln(({1-\kappa})/({1-\zeta}) $ in \eqref{6.3}.\\
Now recall the definitions \eqref{4.17} and \eqref{4.18} and set
$\nu^{\kappa}=(\wt{y}^{\kappa}_\zs{t}\,,\,v^{\kappa}_\zs{t})_\zs{0\le t\le T}$.
Thus for $\nu\in \cU$ with
$V_\zs{T}=-\ln(1-\kappa)$ we have
$$
J(x,\nu)\,\le  J(x,\nu^\kappa)= A(x)+\Gamma(\kappa)\,.
$$
It is clear that \eqref{4.21} gives the optimal value for the parameter $\kappa$. \\
To finish the proof
we have to verify condition \eqref{6.2} for the strategy
$\nu^{*}$ defined in \eqref{4.25}. Indeed,  we have
\begin{align*}
L_\zs{t}(\nu^{*})&=(y^{*},\theta)_\zs{t}-\frac{1}{2}\|y^{*}\|^2_\zs{t}-
|q_\zs{\alpha}|\,\|y^{*}\|_\zs{t}-\int^{t}_\zs{0}\,v^{*}_\zs{s}\d s\\
&=: -\int^{t}_\zs{0}\,g(u)\,\d u\,-\int^{t}_\zs{0}\,v^{*}_\zs{s}\d s\,,
\end{align*}
where
$$
g(t)=\tau^{*}_\zs{t}|\theta_\zs{t}|^{2}
\left(
|q_\zs{\alpha}|\,\chi(t)
-1+
\frac{\tau^{*}_\zs{t}}{2}
\right)
\quad\mbox{and}\quad
\chi(t)=\frac{\tau^{*}_\zs{t}}
{2\sqrt{\int^{t}_\zs{0}(\tau^{*}_\zs{s})^{2}|\theta_\zs{s}|^2\,\d s}}\,.
$$
We recall $\phi(\kappa)$ from \eqref{4.16-2}
and $\kappa_1$ from \eqref{4.21}, then
$$
\tau^{*}_\zs{t}=\tau_{\upsilon_\zs{1}}(t)
\quad
\mbox{with}
\quad
\upsilon_\zs{1}=\phi(\gamma)
\,.
$$
Definition \eqref{4.13} implies
$$
\chi(t)\ge
\frac{\tau_{\upsilon_\zs{1}}(T)}
{2\tau_{\upsilon_\zs{1}}(0)\|\theta\|_\zs{T}}\ge
\frac{1+\upsilon_\zs{1}}{2\|\theta\|_\zs{T}\,
(1+T+\upsilon_\zs{1})}
\ge
\frac{1}{2\|\theta\|_\zs{T}\,(1+T)}\,.
$$
Therefore, condition \eqref{4.23} guarantees that
$g(t)\ge 0$ for $t\ge 0$, which implies
$$
L_\zs{t}(\nu^{*})\ge L_\zs{T}(\nu^{*})=\ln(1-\zeta)\,.
$$
This concludes the proof of Theorem~\ref{Th.4.3}.
\halmos

\medskip

\noindent{\bf Proof of Corollary~\ref{Co.4.4}. \,}
Consider now the optimization problem \eqref{4.21}.
To solve it we have to find the
derivative of the integral in \eqref{4.20}
$$
E(\kappa):= \int^{T}_\zs{0}\omega(t)\,|\theta_\zs{t}|^{2}\,
\left(
{\tau}_{\phi(\kappa)}(t)-\frac12 {\tau}_{\phi(\kappa)}^{2}(t)
\right)\,\d t\,.$$
Indeed, we have with $\phi(\kappa)$ as in \eqref{4.16-2},
\begin{align*}
\dot E(\kappa)=\int^{T}_\zs{0}\omega(t)|\theta_\zs{t}|^{2}\,
\left(
1-\tau_{\phi(\kappa)}(t)
\right)\,
\frac{\partial }{\partial \kappa}\,\tau_{\phi(\kappa)}(t)\,
\d t\,.
\end{align*}
Defining $\tau_\zs{1}(t,{\phi}(\kappa))
:=\frac{\partial \tau_\la(t)}{\partial \la}|_{\la=\phi(\kappa)}$
 we obtain
\begin{equation}\label{6.6}
\frac{\partial }{\partial \kappa}\,\tau_{\phi(\kappa)}(t) =
\tau_\zs{1}(t,{\phi}(\kappa))\,
\frac{\d}{\d\kappa}\,{\phi}(\kappa)\,.
\end{equation}
Therefore,
$$
\dot E(\kappa)=-\frac{1}{1-\kappa}\,B({\phi}(\kappa))
$$
with
$$
B(\lambda)=\frac{\int^{T}_\zs{0}\omega(t)|\theta_\zs{t}|^{2}\,
\left(
1-\tau_\la(t)
\right)\,\tau_\zs{1}(t,\lambda)
\,
\d t}{\dot{\Phi}(\lambda)}\,.
$$
Define $\wh{\tau}(t,\lambda):=|q_\al|\tau_\la(t)/\|\tau_\la\theta\|_T$.
Then, in view of Lemma~\ref{Le.A.3}, we have $\tau_\zs{1}(t,\lambda)\le0$
and, therefore, taking representation \eqref{A.11} into account we obtain
$$
B(\lambda)=\frac{\int^{T}_\zs{0}\omega(t)|\theta_\zs{t}|^{2}\,
\left(
1-\tau(t,\lambda)
\right)\,|\tau_\zs{1}(t,\lambda)|
\,\d t}{\int^{T}_\zs{0}\,\wh{\tau}(t,\lambda)\,
|\tau_\zs{1}(t,\lambda)|\,
\d t}\,.
$$
Moreover, using the lower bound \eqref{A.12} we  estimate
\begin{equation}\label{6.6-1}
B(\lambda)<
\frac{(1+T)^{2}|\theta|^{2}_\zs{\infty}\,\|\theta\|_\zs{T}}
{|q_\zs{\alpha}|-(T+1)\,\|\theta\|_\zs{T}}=:B_{\max}\,.
\end{equation}
Condition \eqref{4.27} for $0< \zeta <\kappa_\zs{0}$ implies that
$$
B_\zs{\max}\le
\left(
\frac{1}{\zeta}-1
\right)\,T-1\,.
$$
Thus for $0\le \kappa \le \zeta <\kappa_\zs{0}$ we obtain
\begin{align*}
\dot \Gamma(\kappa)
\,>\,
\frac{T}{\kappa}-
\frac{1}{1-\kappa}\left(
1+B_\zs{\max}
\right)
\,\ge\,
\frac{T}{\zeta}-
\frac{1}{1-\zeta}\left(
1+B_{\max}
\right)\,\ge\,0
\,.
\end{align*}
This implies $\gamma=\zeta$ and, therefore,
${a}(\gamma):=\ln(({1-\gamma})/({1-\zeta})=0$,
which implies also by Lemma~\ref{Le.4.2-1} that $\phi({a}(\gamma))=\lambda_\zs{\max}$.
Therefore,  we conclude from \eqref{6.4} that
$y^{*}_\zs{t}=\tau_{\lambda_\zs{\max}}(t)\theta_\zs{t}=0$ for all $0\le t\le T$.
\halmos

\medskip

\noindent{\bf Proof of Theorem~\ref{Th.4.5}. \,}
It suffices to verify condition \eqref{6.2} for the strategy
$\nu^*=(y^{*}_\zs{t}\,,\,v^{*}_\zs{t})_\zs{0\le t\le T}$ with
$y^*_\zs{t}=\theta_\zs{t}$ and $v^{*}_\zs{t}=1/\omega(t)$ for
$t\in [0,T]$.
It is easy to show that condition
\eqref{4.28} implies that
$L_\zs{T}(\nu^{*})\ge \ln(1-\zeta)$.
Moreover, for $0\le t\le T$ we can represent
$L_\zs{t}(\nu^{*})$ as
$$
L_\zs{t}(\nu^{*})=-\int^{t}_\zs{0}\,g^{*}_\zs{s}\d s-
\int^{t}_\zs{0}\,v^{*}_\zs{s}\,\d s\,,
$$
where
$$
g^{*}_\zs{t}=
\left(
\frac{|q_\zs{\alpha}|}{\|\theta\|_\zs{t}}
-1
\right)\,\frac{|\theta_\zs{t}|^{2}}{2}
\ge \left(
\frac{|q_\zs{\alpha}|}{\|\theta\|_\zs{T}}
-1
\right)\,\frac{|\theta_\zs{t}|^{2}}{2}\ge 0
$$
since we have assumed that $|q_\al|\ge\|\theta\|_T$.
Therefore,  $L_\zs{t}(\nu^{*})$ is decreasing in $t$;
i.e. $L_\zs{t}(\nu^{*})\ge L_\zs{T}(\nu^{*})$ for all $0\le t\le T$.
This implies the assertion of Theorem~\ref{Th.4.5}.
\halmos

\section{Results and proofs of Section~\ref{subsec:4.3}}

Next we introduce the constraint
\begin{equation}\label{5.2}
K_\zs{1}(y):=
-(y,\theta)_\zs{T}
-f_\zs{\alpha}\left(\|y\|_\zs{T}\right)
\end{equation}
with $f_\alpha(x):=\ln F_\zs{\alpha}(|q_\al|+x)$
and $F_\zs{\alpha}$ introduced in \eqref{Falpha}.

For  $0<a\le -\ln (1-\zeta)$ we consider the following optimization problems
\begin{equation}\label{5.4}
\max_\zs{y\in \cL_\zs{2}[0,T]}\,H(y)
\quad\mbox{subject to}\quad
K_\zs{1}(y)=a\,.
\end{equation}

The following result is the analog of Proposition~\ref{prop1}.

\begin{proposition}\label{prop2}
Assume that the conditions of Lemma~\ref{Le.4.7-1} hold.
Then the optimization problem \eqref{5.4} has the unique solution
$y^*_\zs{t}={\wt y}^{1,a}_\zs{t}=\theta_t \varsigma_{\lambda_\zs{1,a}}(t)$
with $\lambda_\zs{1,a}=\Phi^{-1}_\zs{1}(a)$.
\end{proposition}

\noindent{\bf Proof. \,}
As in the proof of Proposition~\ref{prop1} we use Lagrange's method.
We consider the unconstrained problem
\begin{equation}\label{5.9}
\max_\zs{y\in \cL_\zs{2}[0,T]}\,\Psi_\zs{1}(y,\lambda)\,,
\end{equation}
where $\Psi_\zs{1}(y,\lambda)=H(y)-\lambda K_\zs{1}(y)$
and $\lambda\ge0$ is the Lagrange multiplier.
Taking into account the defining $f_\zs{\alpha}$ in \eqref{5.2}, we obtain the representation
$$
\Psi_\zs{1}(y,\lambda)=\int^{T}_\zs{0}
\left(
\,(\omega(t)+\lambda\,)\,\theta'_\zs{t}\,y_\zs{t}\,
-\frac{\omega(t)}{2}\,|y_\zs{t}|^2
\right)\,
\d t+\lambda\,
f_\zs{\alpha}\left(\|y\|_\zs{T}\right)\,.
$$
Its Fr\'echet derivative is given by
$$
D_\zs{1,y}(h,\lambda)=\lim_\zs{\delta\to 0}\,
\frac{\Psi_\zs{1}(y+\delta h,\lambda)-\Psi_\zs{1}(y,\lambda)}{\delta}\,.
$$
It is easy to show directly that for $\|y\|_\zs{T}>0$
$$
D_\zs{1,y}(h,\lambda)=\int^{T}_\zs{0}\,
(d_\zs{1,y}(t,\lambda))'h_\zs{t}\,\d t\,,
$$
where
$$
d_\zs{1,y}(t,\lambda)=(\omega(t)+\lambda)\theta_\zs{t}-
\omega(t)\,y_\zs{t}
+\lambda \dot{f}_\zs{\alpha}(\|y\|_\zs{T})
\,\ov{y}_\zs{t}\,,
$$
and $\dot{f}_\zs{\alpha}(\cdot)$ denotes the derivative of $f_\zs{\alpha}(\cdot)$.\\
If $\|y\|_\zs{T}=0$, then
$$
D_\zs{1,y}(h,\lambda)=\int^{T}_\zs{0}\,
(\omega(t)+\lambda)\,\theta'_\zs{t}\,h_\zs{t}\,\d t
+\lambda\,\dot{f}_\zs{\alpha}(0)\|h\|_\zs{T}
\,.
$$
We set now
\begin{equation}\label{5.10}
\Delta_\zs{1,y}(h,\lambda)=
\Psi_\zs{1}(y+h,\lambda)-\Psi_\zs{1}(y,\lambda)-D_\zs{1,y}(h,\lambda)\,,
\end{equation}
and show that $\Delta_\zs{1,y}(h,\lambda)\le 0$ for all $y,h\in\cL_\zs{2}[0,T]$. Indeed, if $\|y\|_\zs{T}=0$, then
$$
\Delta_\zs{1,y}(h,\lambda)=-\frac{1}{2}\,\int^{T}_\zs{0}\,\omega(t)\,
|h_\zs{t}|^2\,\d t\,\le\,0\,.
$$
Let now $\|y\|_\zs{T}>0$ and $\ov y=y/\|y\|_T$. Then
$$
\Delta_\zs{1,y}(h,\lambda)=-\frac{1}{2}\,\int^{T}_\zs{0}\,\omega(t)\,
|h_\zs{t}|^2\,\d t\,+\lambda\,\delta_\zs{1,y}(h)\,,
$$
where
$$
\delta_\zs{1,y}(h)=f_\zs{\alpha}(\|y+h\|_\zs{T})
-f_\zs{\alpha}(\|y\|_\zs{T})-
\dot{f}_\zs{\alpha}(\|y\|_\zs{T})\,(\ov{y},h)_\zs{T}\,.
$$
Moreover, by Taylor's formula and denoting by $\ddot{f}_\zs{\alpha}$ the second derivative of $f_\zs{\alpha}$, we get
$$
\delta_\zs{1,y}(h)=
\dot{f}_\zs{\alpha}(\|y\|_\zs{T})\,l_\zs{y}(h)+
\frac{1}{2}\,\ddot{f}_\zs{\alpha}\left(\vartheta\right)
\left(\|y+h\|_\zs{T}
-\|y\|_\zs{T}\right)^2\,,
$$
where  $l_\zs{y}(\cdot)$ is defined in \eqref{5.7} and
$$
\min(\|y\|_\zs{T}\,,\, \|y+h\|_\zs{T})
\le\vartheta\le
\max(\|y\|_\zs{T}\,,\, \|y+h\|_\zs{T})\,.
$$
Recalling the definition of $\varpi$ in \eqref{4.32}, the derivatives of $f_\zs{\alpha}$ are given by
\begin{equation}\label{5.11}
\dot{f}_\zs{\alpha}(x)=-\frac{1}{\varpi(x_\zs{1})}
\quad\mbox{and}\quad
\ddot{f}_\zs{\alpha}(x)=-\frac{1-x_\zs{1}\,\varpi(x_\zs{1})}{\varpi^2(x_\zs{1})}\,.
\end{equation}
The right inequality in \eqref{4.33} and
Lemma~\ref{Le.A.2} imply that $\Delta_\zs{1,y}(h,\lambda)\le 0$
for all $\lambda\ge 0$ and $y,h\in\cL_\zs{2}[0,T]$.
The solution of the optimization problem \eqref{5.9} is given by
$y\in\cL_\zs{2}[0,T]$ such that
\begin{equation}\label{5.12}
D_\zs{1,y}(h,\lambda)=0
\quad\mbox{ for all }\quad
h\in\cL_\zs{2}[0,T]\,.
\end{equation}
Notice that for $\|\theta\|_\zs{T}>0$ the solution  \eqref{5.12} can not be zero, since
for $y=0$ we obtain  $D_\zs{1,y}(h,\lambda)<0$ for $h=-\theta$.
Therefore, we have to solve equation \eqref{5.12} for $y$ with $\|y\|_\zs{T}>0$, equivalently, we have to find a non-zero function in $\cL_\zs{2}[0,T]$ satisfying
$$
d_\zs{1,y}(t,\lambda)=0\,.
$$
One can show directly that for $0\le \lambda\le \lambda^*_\zs{\max}$
the solution of this equation is given by
\begin{equation}\label{5.12-1}
y^{1,\lambda}_\zs{t}=\varsigma_\la(t)\theta_\zs{t}\,,
\end{equation}
where $\varsigma_\la(t)$ is defined in  \eqref{4.37}.
Now we have to choose the parameter $\lambda$ to satisfy the constraint in \eqref{5.4}.
Note that
$$
K_\zs{1}(y^{1,\lambda})=\Phi_\zs{1}(\lambda)\,.
$$
Under the conditions of Lemma~\ref{Le.4.7-1} the inverse of $\Phi_\zs{1}$ exists.
Therefore, the function $\ov y^{\la_\zs{a}}\not\equiv 0$ with $\lambda_\zs{a}=\Phi_\zs{1}^{-1}(a)$
is the solution of the optimization problem \eqref{5.4}.
\halmos

\medskip

\noindent{\bf Proof of Theorem~\ref{Th.4.8}.\,}
Define
\beam\label{L}
\ov L_\zs{t}(\nu)=
(y,\theta)_\zs{t}\,-\,V_\zs{t}\,+\,
f_\zs{\alpha}\left(\|y\|_\zs{t}\right)\,,\quad 0\le t\le T\,,
\eeam
with $f_\zs{\alpha}$ defined in \eqref{4.39}.\\
First note that the risk bound in the optimization problem \eqref{4.30} is equivalent to
\begin{equation}\label{6.7}
\inf_\zs{0\le t\le T}\,\ov L_\zs{t}(\nu)\,\ge\,
\ln\,\left(1-\zeta\right)\,,
\end{equation}
As in the proof of Theorem~\ref{Th.4.3}
we start with the constraint at time $t=T$:
$$
\ov L_\zs{T}(\nu)\,\ge\, \ln\,\left(1-\zeta\right)\,.
$$
Taking the definition of $K_\zs{1}$ in \eqref{5.2} into account and choosing $V=V^\kappa$ as in \eqref{6.1} we rewrite this inequality as
$$
K_\zs{1}(y)\, \le\,\ln\frac{1-\kappa}{1-\zeta}\,,\quad 0\le \kappa\le \zeta\,.
$$
To find the optimal strategy we use the optimization problem \eqref{5.4},
extending the range of $a$ to  $0\le a\le \ln(({1-\kappa})/({1-\zeta})$.
In Proposition~\ref{prop2} we established that
for each $0<a\le -\ln(1-\zeta)$
\begin{equation}\label{6.7-1}
\max_\zs{y\in \cL_\zs{2}[0,T]\,,\,K_\zs{1}(y)=a}\,H(y)\,=
H(\ov y^{\Phi_\zs{1}^{-1}(a)})=:\ov C(\Phi_\zs{1}^{-1}(a))\,,
\end{equation}
where $y^{1,\lambda}$ is defined in \eqref{5.12-1} and
$$
\ov C(\lambda)=\int^{T}_\zs{0}\,\omega(t)|\theta_\zs{t}|^2
\left(
\varsigma_\la(t)-\frac{1}{2}\varsigma^2_\la(t)
\right)\,
\d t\,.
$$
To study the optimization problem \eqref{5.4} for $a=0$ note that
$$
K_\zs{1}(y)\ge k_\zs{\min}(\|y\|_\zs{T})
\quad
\mbox{with}
\quad
k_\zs{\min}(x)=
-x\|\theta\|_\zs{T}
-f_\zs{\alpha}(x)\,,quad x\ge 0\,.
$$
Moreover,
$$
\dot{k}_\zs{\min}(x) =\frac{1}{\varpi(|q_\zs{\alpha}|+x)}-|\theta\|_\zs{T}\,,quad x\ge 0\,,
$$
and by the right inequality in \eqref{4.33} we obtain
for $|q_\zs{\alpha}|>\|\theta\|_\zs{T}$ (which follows from
condition \eqref{4.23})
$$\
\dot{k}_\zs{\min}(x)\ge |q_\zs{\alpha}|+x-|\theta\|_\zs{T}>0\,,quad x\ge 0,.
$$
Therefore, $k_\zs{\min}(x)>k_\zs{\min}(0)=0$ for all $x>0$
and $k_\zs{\min}(x)=0$ if and only if $x=0$.
This means that only $y\equiv 0$ satisfies  $K_\zs{1}(y)=0$.
Moreover, in view of Lemma~\ref{Le.4.7} and Lemma~\ref{Le.4.7-1},
 as in the proof of Theorem~\ref{Th.4.3}, we obtain
$\ov {\wt y}^{0}=0$.
Therefore, the function $\ov {\wt y}^{)}$ is the solution of \eqref{5.4} for all $0\le a\le -\ln(1-\zeta)$.\\
To choose the parameter $0\le a\le \ln(({1-\kappa})/({1-\zeta})$ we  calculate
the derivative of $\ov C(\Phi_\zs{1}^{-1}(a))$ as
$$
\frac{\d }{\d a} {\ov C}(\Phi_\zs{1}^{-1}(a))
= \frac{\d }{\d a}{\Phi_\zs{1}}^{-1}(a)
\int^{T}_\zs{0}\,\omega(t)|\theta_\zs{t}|^2\,
\left(
1-\varsigma_{\Phi_\zs{1}^{-1}(a)}(t)
\right)\,
\frac{partial}{\partial\la}\varsigma_\la(t)|_{\la=\Phi_\zs{1}^{-1}(a)}\,,
\d t\,.
$$
We recall that
$$
\frac{\d}{\d a}\,\Phi_\zs{1}^{-1}(a)=\frac{1}{\dot{\Phi_\zs{1}}(\Phi_\zs{1}^{-1}(a))}
\quad\mbox{with}\quad
\Psi_\zs{1}(a)=\frac{\d}{\d a}\Phi_\zs{1}(a)\,.
$$
Therefore, by Lemma~\ref{Le.A.4}, the derivative of
$\frac{\d }{\d a}\ov C(\Phi_\zs{1}^{-1}(a))>0$, which implies that
$$
\max_{\zs{0}\le a\le \ln(({1-\kappa})/({1-\zeta})}\,
\ov C(\Phi_\zs{1}^{-1}(a))\,=
\ov C(\Phi_\zs{1}^{-1}\left(\ln\frac{1-\kappa}{1-\zeta}\right)\,.
$$
So in \eqref{6.7-1} we take $a=\ln(({1-\kappa})/({1-\zeta})$.\\
Recalling the notation $\ov{\wt y}^\kappa=\varsigma_{\Phi_\zs{1}(\kappa)}(t)$ from \eqref{3.30a} we set
$\ov{\nu}^{\kappa}=(\ov{\wt y}^{\kappa}_\zs{t}\,,\,v^{\kappa}_\zs{t})_\zs{0\le t\le T}$.
Then, for $\nu\in \cU$ with $V_\zs{T}=-\ln(1-\kappa)$,
$$
J(x,\nu)\,\le  J(x,\ov{\nu}^\kappa)= A(x)+\Gamma_\zs{1}(\kappa)\,.
$$
It is clear that \eqref{4.43} gives the optimal value for the parameter
$\kappa$.\\
To finish the proof
we have to verify condition \eqref{6.7} for the strategy
$\nu^{*}$ as defined in \eqref{4.47}.
To this end, with $\phi(\kappa)=\Phi_\zs{1}^{-1}\big(\ln((1-\kappa)/(1-\zeta))\big)$, we set
$$
\varsigma^*_t =\varsigma_{\phi_\zs{1}}(t)\,,
\quad
\phi_\zs{1}=
\phi_\zs{1}(\gamma_\zs{1})
\quad\mbox{and}\quad
\gamma^*(t)=\frac{\varsigma^*_\zs{t}}{2\|\varsigma^*\theta\|_\zs{t}}\,.
$$
With this notation we can represent the function
$L_\zs{t}(\nu^{*})$ in the following
 integral form
\begin{align*}
L_\zs{t}(\nu^{*})
=-\int^{t}_\zs{0}\,g^*(u)\,\d u\,-\int^{t}_\zs{0}\,v^{*}_\zs{s}\d s\,,
\end{align*}
where
$$
g^*(t)=\varsigma^{*}_\zs{t}|\theta_\zs{t}|^{2}
\left(
\frac{\beta_\zs{t}\gamma^*(t)}{2}\,
\,
-1
\right)
\quad\mbox{with}\quad
\beta_\zs{t}=-
\dot{f}_\zs{\alpha}\left(\|\varsigma^*\theta\|_\zs{t}\right)\,.
$$
Note that definition \eqref{4.37} and the inequalities \eqref{4.38} imply
$$
\gamma^*(t)\ge
\frac{\varsigma_{\wh{\phi}_\zs{1}}(T)} {2\varsigma_{\phi_\zs{1}}(0)\|\theta\|_\zs{t}}
\ge
\frac{1+\phi_\zs{1}}{2\|\theta\|_\zs{t}\,(1+T+\phi_\zs{1})}
\ge
\frac{1}{2\|\theta\|_\zs{T}\,(1+T)}\,.
$$
Moreover, from the right inequality in \eqref{4.33}
we obtain
$$
\beta_\zs{t}=\frac{1}{\varpi\left(\|\varsigma^*\theta\|_\zs{t}\right)}
\ge |q_\zs{\alpha}|+\|\varsigma^*\theta\|_\zs{t}\ge |q_\zs{\alpha}|\,.
$$
Therefore, condition \eqref{4.23} implies that
$g^*(t)\ge 0$, i.e.
$$
L^*_\zs{t}(\nu^{*})\ge L^*_\zs{T}(\nu^{*})=\ln(1-\zeta)\,.
$$
This concludes the proof of Theorem~\ref{Th.4.8}.
\halmos

\medskip

\noindent{\bf Proof of Corollary~\ref{Co.4.9}. \,}
Consider now the  optimization problem \eqref{4.43}.
To solve this we have to calculate the
derivative of $\ov E$ in \eqref{4.42}.
We obtain
\begin{align*}
\frac{\d}{\d \kappa}
\ov E(\kappa)=\int^{T}_\zs{0}\omega(t)|\theta_\zs{t}|^{2}\,
\left(
1-{\varsigma}_{}(t)
\right)\,
\frac{\partial }{\partial \kappa}\,\wh{\varsigma}(t,\kappa)\,
\d t\,.
\end{align*}
We recall from \eqref{4.41} that $\phi_\zs{1}(\kappa)=\Phi_\zs{1}^{-1}(\ln\frac{1-\kappa}{1-\zeta})$
and define the partial derivative
$\varsigma_\zs{1}(t,\lambda)=\frac{\partial }{\partial\la}\varsigma_\la(t)$.
Then
\begin{equation}\label{6.11}
\frac{\partial }{\partial \kappa}\,{\varsigma}(t,\kappa)=
\varsigma_\zs{1}(t,\phi_\zs{1}(\kappa))\,
\frac{\d }{\d\kappa}\,\phi_\zs{1}(\kappa)\,.
\end{equation}
Therefore,
$$
\frac{\d}{\d \kappa} \ov E(\kappa)
=-\frac{1}{1-\kappa}\,\ov B(\Phi_\zs{1}^{-1}(\kappa))
$$
with
$$
\ov B(\lambda)=\frac{\int^{T}_\zs{0}\omega(t)|\theta_\zs{t}|^{2}\,
\left(
1-\varsigma_\la(t)
\right)\,\varsigma_\zs{1}(t,\lambda)
\,
\d t}{\dot{\Phi}(\lambda)}\,.
$$
By Lemma~\ref{Le.A.4},   $\varsigma_\zs{1}(t,\lambda)\le 0$,
therefore, taking  representation
\eqref{A.20} into account, we obtain
$$
\ov B(\lambda)=\frac{\int^{T}_\zs{0}\omega(t)|\theta_\zs{t}|^{2}\,
\left(
1-\varsigma_\la (t)
\right)\,|\varsigma_\zs{1}(t,\lambda)|
\,\d t}{\int^{T}_\zs{0}\,\eta(t,\lambda)\,
|\varsigma_\zs{1}(t,\lambda)|\,
\d t}\,.
$$
Moreover, with the lower bound \eqref{A.21}
we can estimate $\ov B(\lambda)$  as in in \eqref{6.6-1}, i.e.
$$
\ov B(\lambda)\le B_\zs{\max}\,.
$$
The remainding proof is the same as  the proof
 of Corollary~\ref{Co.4.9}.
\halmos

\medskip

\noindent{\bf Proof of Theorem~\ref{Th.4.10}. \,}
We have to verify condition \eqref{6.7} for the strategy
$\nu^*=(y^{*}_\zs{t}\,,\,v^{*}_\zs{t})_\zs{0\le t\le T}$ with
$y^{*}_\zs{t}=\theta_\zs{t}$ and $v^{*}_\zs{t}=1/\omega(t)$ for
$t\in [0,T]$. \\
First note that condition \eqref{4.49} implies
$$
\ov L_\zs{T}(\nu^{*})\ge \ln(1-\zeta)\,.
$$
Moreover, for $0\le t\le T$ we can represent the function
$\ov L_\zs{t}(\nu^{*})$ as
$$
\ov L_\zs{t}(\nu^{*})=\|\theta\|^2_\zs{t}
+f_\zs{\alpha}(\|\theta\|_\zs{t})-V^*_\zs{t}
=-\int^{t}_\zs{0}\,l^{*}_\zs{s}\d s-
\int^{t}_\zs{0}\,v^{*}_\zs{s}\,\d s\,,
$$
where
$$
l^{*}_\zs{t}=
\left(
\frac{1}{\varpi(|q_\zs{\alpha}|+\|\theta\|_\zs{t})}
-1
\right)\,|\theta_\zs{t}|^{2}\,.
$$
Therefore, by the right inequality in \eqref{4.33}
we obtain
$$
l^{*}_\zs{t}\ge \left(
|q_\zs{\alpha}|+
\|\theta\|_\zs{t}
-1
\right)\,|\theta_\zs{t}|^{2}
\ge
\left(
|q_\zs{\alpha}|
-1
\right)\,|\theta_\zs{t}|^{2}
$$
and by condition \eqref{4.45} we get $l^*_\zs{t}> 0$
 for $0\le t\le T$, therefore,  $\ov L_\zs{t}(\nu^{*})$ is decreasing in $t$, i.e. for $0<t\le T$
$$
\ov L_\zs{t}(\nu^{*})\ge \ov L_\zs{T}(\nu^{*})\ge \ln(1-\zeta)\,.
$$
This concludes the proof of Theorem~\ref{Th.4.10}.
\halmos

\renewcommand{\theequation}{A.\arabic{equation}}
\renewcommand{\thetheorem}{A.\arabic{theorem}}
\renewcommand{\thesection}{A.\arabic{section}}
\chapter{Appendix}

\section{Results for Section~\ref{subsec:4.2}}\label{subsec:A.1}

{\bf Proof of Lemma~\ref{Le.4.2}. \,}
Since $G(u,\la)$ is for fixed $\la$ decreasing to 0 in $u$,
equation $G(u,\lambda)=1$
has a positive solution if and only if $G(0,\lambda)\ge 1$.
But this is equivalent to $k_2+2\la k_1-\la^2(|q_\al|^2-\|\theta|\_T^2)\ge 0$,
which gives the upper bound for $\la$.
Moreover, taking into account that $G(0,\lambda_\zs{\max})=1$
we obtain through the definition \eqref{4.11} $\rho(\lambda_\zs{\max})=0$
\halmos

\medskip

Next we prove some properties of  $\Phi$ and $\tau_{\phi(\kappa)}$.

\begin{lemma}\label{Le.A.3}
The function $\tau_\la(t)$ is continuously differentiable in $\la$ for $0\le \lambda\le \lambda_\zs{\max}$ with partial derivative
$$\tau_\zs{1}(t,\lambda)=\frac{\partial}{\partial\la}\tau_\la(t) <0\,,\quad 0\le t\le T\,.$$
Moreover, under the condition \eqref{4.23}
the derivative  $\dot{\Phi}(\lambda)< 0$ for
$0\le \lambda\le\lambda_\zs{\max}$.
\end{lemma}

\noindent {\bf Proof. \,}
First note that
$$
\tau_\zs{1}(t,\lambda)=
-
|q_\zs{\alpha}|
\frac{
\left(\rho(\lambda)\omega(t)
-\lambda\dot{\rho}(\lambda)(\omega(t)+\lambda)
\right)
}
{\left(\lambda |q_\zs{\alpha}| +
\rho(\lambda)(\omega(t)+\lambda)\right)^2}\,.
$$
By the definition of $\rho(\lambda)$ in \eqref{4.11} we get
$G(\rho(\lambda),\lambda)=1$ for $0\le \lambda\le \lambda_\zs{\max}$.
Therefore,
$$
\dot{\rho}(\lambda)=-\frac{G_\zs{2}(\rho(\lambda),\lambda)}
{G_\zs{1}(\rho(\lambda),\lambda)}
$$
with
$$
G_\zs{1}(u,\lambda)=\frac{\partial G(u,\lambda)}{\partial u}
\quad\mbox{and}\quad
G_\zs{2}(u,\lambda)=\frac{\partial G(u,\lambda)}{\partial \lambda}\,.
$$
The definition of $G$ in \eqref{4.10} implies that
$$
G_\zs{1}(u,\lambda)=-2\int^{T}_\zs{0}
\frac{(\omega(t)+\lambda)^{3}}{(\lambda|q_\zs{\alpha}|+u(\omega(t)+\lambda))^{3}}
\,|\theta_\zs{t}|^2\,\d t
$$
and
$$
G_\zs{2}(u,\lambda)=-2|q_\zs{\alpha}|\int^{T}_\zs{0}
\frac{\omega(t)(\omega(t)+\lambda)}
{(\lambda|q_\zs{\alpha}|+u(\omega(t)+\lambda))^{3}}\,
|\theta_\zs{t}|^2
\,\d t\,.
$$
Therefore, for all
$0\le \lambda\le \lambda_\zs{\max}$
and
$0\le t\le T$
$$
\dot{\rho}(\lambda)< 0
\quad\mbox{and}\quad
\tau_\zs{1}(t,\lambda)< 0\,.
$$
We calculate now the derivative of $\Phi$ as
\begin{equation}\label{A.11}
\dot{\Phi}(\lambda)=
\int^{T}_\zs{0}
\,\wh{\tau}(t,\lambda)\,
\tau_\zs{1}(t,\lambda)\,|\theta_\zs{t}|^2
\d t\,,
\end{equation}
where
$$
\wh{\tau}(t,\lambda)=
\frac{|q_\zs{\alpha}|\tau(t,\lambda)}{\|\tau_\zs{\lambda}\theta\|_\zs{T}}
-1+\tau(t,\lambda)\,.
$$
To estimate this term from below note that by the
inequlities \eqref{4.14}
$$
\frac{\tau(t,\lambda)}{\|\tau_\zs{\lambda}\theta\|_\zs{T}}\,
\ge \,
\frac{\tau(T,\lambda)}{\tau(0,\lambda)\|\theta\|_\zs{T}}\ge
 \frac{1}{(T+1)\|\theta\|_\zs{T}}\,.
$$
Therefore,
\begin{equation}\label{A.12}
\wh{\tau}(t,\lambda)\ge \frac{|q_\zs{\alpha}|}{(T+1)\|\theta\|_\zs{T}}
-1
\end{equation}
and by the condition \eqref{4.23}
$\wh{\tau}(t,\lambda)> 0$ for $0\le t\le T$ and
$0\le \lambda\le \lambda_\zs{\max}$, i.e.
$\dot{\Phi}(\lambda)<0$.
\halmos

\medskip

\noindent{\bf Proof of Lemma~\ref{Le.4.2-1}.\,}
Taking into account that $\tau_0(\cdot)\equiv 1$ we get
$$
\Phi(0)=|q_\zs{\alpha}|\|\theta\|_\zs{T}-\frac{1}{2}\|\theta\|^2_\zs{T}\,.
$$
Moreover, condition \eqref{4.16-1}
implies $\Phi(0)>-\ln(1-\zeta)$. Therefore, in view of \eqref{6.4}
and Lemma~\ref{Le.A.3} we get that the inverse  $\Phi^{-1}(a)$
exists for $0< a\le -\ln(1-\zeta)$
with $0\le\Phi^{-1}(a)<\lambda_\zs{\max}$ and
$\Phi^{-1}(0)=\lambda_\zs{\max}$.
\halmos

\section{Results for Section~\ref{subsec:4.3}}\label{subsec:A.2}

We present some properties of  $\Phi_\zs{1}(\lambda)$ and $\varsigma_{\phi(\ov\kappa)}$.

\begin{lemma}\label{Le.A.4}
The function $\varsigma_\la(t)$ is continuously differentiable in $\la$ for all
$0\le \lambda\le \lambda^*_\zs{\max}$
with partial derivative
$$\varsigma_\zs{1}(t,\lambda)=\frac{\partial }{\partial\la} \varsigma_\la(t)<0\,,\quad \le t\le T\,.$$
Moreover, under condition \eqref{4.23}
the derivative  $\dot{\Phi_\zs{1}}(\lambda)< 0$ for
$0\le \lambda\le\lambda^*_\zs{\max}$.
\end{lemma}

\noindent {\bf Proof. \,} First note that
\begin{equation}\label{A.17}
\varsigma_\zs{1}(t,\lambda)=
-
\frac{b(t,\lambda)\,c_\zs{\alpha}(\lambda)}
{\left(b(t,\lambda)+c_\zs{\alpha}(\lambda)\right)^2}
\left(
\frac{\omega(t)}{\lambda(\omega(t)+\lambda)}
-\dot{\rho_\zs{1}}(\lambda)\,\Omega_\zs{\alpha}(\rho_\zs{1}(\lambda))
\right)
\end{equation}
where
$$
\Omega_\zs{\alpha}(\rho_\zs{1})=
\frac{\iota_\zs{\alpha}(\rho_\zs{1})-\rho_\zs{1}\,\dot{\psi}_\zs{\alpha}(\rho_\zs{1})}
{\rho_\zs{1}\,\iota_\zs{\alpha}(\rho_\zs{1})}\,.
$$
Note that we can represent the numerator as
$$
\iota_\zs{\alpha}(\rho_\zs{1})-\rho_\zs{1}\,\dot{\psi}_\zs{\alpha}(\rho_\zs{1})=
\frac{
\varpi(y)\left(
1+y(y-|q_\zs{\alpha}|)
\right)-
(y-|q_\zs{\alpha}|)
}
{
\varpi^2(y)
}
$$
with $y=|q_\zs{\alpha}|+\rho_\zs{1}$.
Therefore, the left inequality in \eqref{4.33} implies
\begin{align*}
\varpi(y)
\left(
1+y(y-|q_\zs{\alpha}|)
\right)-
(y-|q_\zs{\alpha}|)
&\ge
\left(
1+y(y-|q_\zs{\alpha}|)
\right)
\left(
\frac{1}{y}-\frac{1}{y^3}
\right)
-
(y-|q_\zs{\alpha}|)\\
&= \, \frac{y|q_\zs{\alpha}|-1}{y^3}
 \, \ge \,
\frac{q^2_\zs{\alpha}-1}{y^3}
\,,
\end{align*}
and by condition \eqref{4.45}
we obtain
$$
\Omega_\zs{\alpha}(\rho_\zs{1})\ge 0
\quad\mbox{for}\quad
\rho_\zs{1}\ge 0\,.
$$
Let us now calculate $\dot{\rho_\zs{1}}$. To this end note that
 definition \eqref{4.35} implies
$H(\rho_\zs{1}(\lambda),\lambda)=1$ for all $0\le \lambda\le \lambda^*_\zs{\max}$.
Therefore,
$$
\dot{\rho_\zs{1}}(\lambda)=-\frac{H_\zs{2}(\rho_\zs{1}(\lambda),\lambda)}
{H_\zs{1}(\rho_\zs{1}(\lambda),\lambda)}
$$
with
$$
H_\zs{1}(u,\lambda)=\frac{\partial H(u,\lambda)}{\partial u}
\quad\mbox{and}\quad
G^*_\zs{2}(u,\lambda)=\frac{\partial G^*(u,\lambda)}{\partial \lambda}\,.
$$
The definition of $H$ in \eqref{4.31} implies that
\begin{equation}\label{A.18}
H_\zs{1}(u,\lambda)=-2\int^{T}_\zs{0}
\frac{(\omega(t)+\lambda)^{2}(\lambda (\dot{\psi}_\zs{\alpha}(u)+1)+\omega(t))}
{(\lambda\,\iota_\zs{\alpha}(u)+u(\omega(t)+\lambda))^{3}}
\,|\theta_\zs{t}|^2\,\d t
\end{equation}
and
\begin{equation}\label{A.19}
H_\zs{2}(u,\lambda)=-2
\iota_\zs{\alpha}(u)
\int^{T}_\zs{0}
\frac{\omega(t)\,(\omega(t)+\lambda)}
{(\lambda\,\iota_\zs{\alpha}(u)+u(\omega(t)+\lambda))^{3}}\,
|\theta_\zs{t}|^2
\,\d t\,.
\end{equation}
Taking into account that
$$
\dot{\psi}_\zs{\alpha}(u)+1=
\frac{1-|q_\zs{\alpha}|+u \varpi(|q_\zs{\alpha}|+u)}{\varpi^2(\varpi(|q_\zs{\alpha}|+u))}\,,
$$
 we obtain from the right inequality in \eqref{4.33}
$$
\dot{\psi}_\zs{\alpha}(x)+1\ge 0
\quad\mbox{for all}\quad
x\ge 0\,.
$$
Therefore, for all
$0\le \lambda\le \lambda^\prime_\zs{\max}$
and
$0\le t\le T$

$$
\dot{\rho_\zs{1}}(\lambda)< 0
\quad\mbox{and}\quad
\varsigma_\zs{1}(t,\lambda)< 0
\,.
$$
Let us calculate now the derivative of $\Phi_\zs{1}$.
We obtain
\begin{equation}\label{A.20}
\frac{\d }{\d \lambda}\,\Phi_\zs{1}(\lambda)=
\int^{T}_\zs{0}
\,\eta(t,\lambda)\,
\varsigma_\zs{1}(t,\lambda)
\d t\,,
\end{equation}
where
$$
\eta(t,\lambda)=
-\frac{\dot{f}_\zs{\alpha}(\|y_\zs{\lambda}\|_T)
\,\varsigma_\la(t)}
{\|y_\zs{\lambda}\|_T}
-1
=
\frac{1}
{\varpi\left(|q_\zs{\alpha}|+\|y_\zs{\lambda}\|_T\right)}
\,\frac{
\varsigma_\la(t)}
{\|y_\zs{\lambda}\|_T}
-1
\,.
$$
with $\|y_\zs{\lambda}\|_T=\|\varsigma_\zs{\lambda}\theta\|_\zs{T}$.
In view of the inequlities \eqref{4.38} we obtain
$$
\frac{
\varsigma_\la(t)}
{\|y_\zs{\lambda}\|_T}=
\frac{\varsigma_\la(t)}{\|\varsigma_\zs{\lambda}\theta\|_\zs{T}}\,
\ge \,
\frac{\varsigma_\la(T)}{\varsigma_\la(0)\|\theta\|_\zs{T}}\ge
 \frac{1}{(T+1)\|\theta\|_\zs{T}}\,.
$$
Therefore, by the right inequality in \eqref{4.33}
and the condition \eqref{4.23}
\begin{equation}\label{A.21}
\eta(t,\lambda)\ge
\frac{|q_\zs{\alpha}|+\|y_\zs{\lambda}\|_T}{(T+1)\|\theta\|_\zs{T}}
-1\,
\ge
\frac{|q_\zs{\alpha}|}{(T+1)\|\theta\|_\zs{T}}
-1>0
\end{equation}
for $0\le t\le T$ and $0\le \lambda\le \lambda^*_\zs{\max}$.
\halmos

\medskip

\noindent{\bf Proof of Lemma~\ref{Le.4.7-1}.\, }
Similarly to the proof of Lemma~\ref{Le.4.2-1}
we observe that condition \eqref{4.40-1} implies
$$
\Phi_\zs{1}(0)=-\|\theta\|_\zs{T}
-\ln F_\zs{\alpha}(|q_\zs{\alpha}|+\|\theta\|_\zs{T})
>-\ln(1-\zeta)\,.
$$
Moreover,
$\Phi_\zs{1}(\lambda^\prime_\zs{\max})=0$ since $\rho_\zs{1}(\lambda^\prime_\zs{\max})=0$.
This means that $\phi^*(0)=\lambda^*_\zs{\max}$.\\
In view of Lemma~\ref{Le.A.4} $\Phi_\zs{1}(\cdot)$ is strictly decraesing
on  $[0,\lambda^\prime_\zs{\max}]$. Therefore, $\Phi_\zs{1}^{-1}$ exists
 for all $0< a\le -\ln(1-\zeta)$
such that $0\le\phi_\zs{1}(a)<\lambda^\prime_\zs{\max}$ with $\phi_\zs{1}(\lambda^\prime_\zs{\max} )=0$.
\halmos

%

%
\bibliographystyle{./wdg_plain}                       %
\bibliography{database}                             %

\end{document}